\documentclass[%
 aip,
 amsmath,amssymb,
 reprint,%
]{revtex4-1}

\usepackage{graphicx}
\usepackage{dcolumn}
\usepackage{bm}

\usepackage[english]{babel}
\usepackage[utf8]{inputenc}
\usepackage[T1]{fontenc}
\usepackage{mathptmx}
\usepackage{etoolbox}
\usepackage{subfig}
\newcommand{\q}{\overline{q}}
\newcommand{\w}{\overline{w}}

\usepackage{xcolor}

\usepackage{hyperref}
\usepackage{cleveref}

\makeatletter
\def\@email#1#2{%
 \endgroup
 \patchcmd{\titleblock@produce}
  {\frontmatter@RRAPformat}
  {\frontmatter@RRAPformat{\produce@RRAP{*#1\href{mailto:#2}{#2}}}\frontmatter@RRAPformat}
  {}{}
}%
\makeatother
\begin{document}

\preprint{AIP/123-QED}



\title[]{The bcc coating of Lennard-Jones crystal nuclei vanishes with a change of local structure detection algorithm}
\author{Willem Gispen}
 \affiliation{ 
Soft Condensed Matter \& Biophysics, Debye Institute for Nanomaterials Science, Utrecht University, Princetonplein 1, 3584 CC Utrecht, Netherlands 
}%

 \author{Alberto Pérez de Alba Ortíz}
\affiliation{Computational Soft Matter Lab, Computational Chemistry Group and Computational Science Lab, van 't Hoff Institute for Molecular Science and Informatics Institute, University of Amsterdam, Science Park 904, 1098 XH Amsterdam, Netherlands.}
\author{Marjolein Dijkstra}%
\affiliation{ 
Soft Condensed Matter \& Biophysics, Debye Institute for Nanomaterials Science, Utrecht University, Princetonplein 1, 3584 CC Utrecht, Netherlands 
}%

\date{\today}

\begin{abstract}
Since the influential work of ten Wolde, Ruiz-Montero, and Frenkel [Phys. Rev. Lett. 75, 2714 (1995)],\cite{ten_wolde_numerical_1995} crystal nucleation from a Lennard-Jones fluid has been regarded as a paradigmatic example of metastable crystal ordering at  the surface of a critical nucleus.
We apply seven commonly used  local structure detection algorithms to characterize crystal nuclei obtained from transition path sampling simulations.
The polymorph composition of these nuclei varies significantly depending on the  algorithm used. 
Our results indicate that one should be very careful when characterizing the local structure near solid-solid and solid-fluid interfaces.
Particles near such interfaces exhibit a local  structure distinct from that of bulk fluid or bulk crystal phases. We argue that incorporating outlier detection into the local structure detection method is beneficial, leading to greater confidence in the classification results. Interestingly, the bcc coating nearly disappears when adopting a machine learning method with outlier detection. 
\end{abstract}



\maketitle


\section{Introduction}
In the simplest view of crystallization, a stable crystal phase forms directly from a supercooled fluid.
However, as early as 1897, Ostwald realized that this simple picture does not always hold: his famous step rule states that the nucleating phase is typically a less stable crystal phase instead. Later, Stranski and Totomanow\cite{stranski_rate_1933} rationalized Ostwald's step rule by proposing that the nucleating phase is determined by the one with the lowest free-energy barrier for nucleation.
Alexander and McTague\cite{alexander_should_1978} further demonstrated that general symmetry considerations uniquely favor the body-centered cubic (bcc) crystal phase near the solid-fluid coexistence. 
These studies all predict an important role for metastable phases during crystal nucleation.
In recent decades, it has become possible to directly test these predictions using computer simulations.
 This has enabled the observation  of intermediate phases in  the crystallization of colloidal fluids,\cite{fortini_crystallization_2008, ji_crystal_2018, kratzer_two-stage_2015,gispen_kinetic_2022} metallic melts,\cite{sadigh_metastablesolid_2021,sun_two-step_2022} and aqueous solutions.\cite{arjun_unbiased_2019,bulutoglu_investigation_2022} These intermediates can be metastable crystal polymorphs,\cite{desgranges_controlling_2007, kratzer_two-stage_2015, sadigh_metastablesolid_2021, gispen_kinetic_2022} amorphous solids,\cite{fortini_crystallization_2008, bulutoglu_investigation_2022} or fluid phases.\cite{fortini_crystallization_2008, van_meel_two-step_2008}

In 1995, \citet{ten_wolde_numerical_1995} identified a remarkable role for bcc during crystal nucleation from a Lennard-Jones fluid which is in simultaneous accordance with the conjectures of Ostwald, Stranski and Totomanow, and Alexander and McTague. They observed that small, precritical nuclei were predominantly bcc-like, while  the critical nucleus was predominantly fcc-like. Surprisingly, the surface of the critical nucleus was also significantly bcc-like. Since the fcc crystal is the stable phase under the conditions they investigated, the initial formation of bcc is in line with Ostwald's step rule. The bcc `coating' is also consistent  with density functional theory calculations, \cite{shen_bcc_1996, wang_density_2013, schoonen_crystal_2022}
which suggest that the presence of bcc lowers the nucleation barrier. 
Since the influential work of \citeauthor{ten_wolde_numerical_1995},\cite{ten_wolde_numerical_1995,moroni_interplay_2005,eslami_local_2017,prestipino_barrier_2018,jungblut_crystallization_2011,ouyang_entire_2020} similar coatings of metastable polymorphs on the surface of crystal nuclei have been identified in water,\cite{russo_new_2014, espinosa_interfacial_2016, prestipino_barrier_2018, leoni_nonclassical_2021} colloids,\cite{kawasaki_formation_2010,lechner_role_2011,tan_visualizing_2014} and metals.\cite{becker_unsupervised_2022}


However, in the case of water, several authors have recently demonstrated that the surface structure is highly dependent on the local structure detection algorithm.\cite{espinosa_interfacial_2016, prestipino_barrier_2018, leoni_nonclassical_2021} Even minor variations in the definition of nearest neighbors 
can lead to the appearance or disappearance of the coating.\cite{espinosa_interfacial_2016} This controversy in water raises the question of whether the bcc coating of Lennard-Jones nuclei exhibits a similar dependence on the local structure detection algorithm.

In this paper, we aim to address  this question by first simulating crystal nucleation from a Lennard-Jones fluid under  the same conditions as those investigated in Ref.\ \citenum{ten_wolde_numerical_1995} using transition path sampling. We characterize the structure of crystal nuclei using seven different local structure detection methods. This includes  two variants of the histogram method from Ref.\ \citenum{ten_wolde_numerical_1995} as well as methods based on more recent advances in local structure detection, such as locally averaged bond-order parameters,\cite{lechner_accurate_2008} machine learning,\cite{chung_data-centric_2022} and polyhedral template matching.\cite{larsen_robust_2016}

\section{Simulation methods}
\label{sec:simulation-details}

\subsection{Molecular dynamics}
We perform molecular dynamics simulations of $N=3\times 10^4$ particles interacting with a Lennard-Jones pair potential
\begin{equation*}
    u(r) = 4 \epsilon \left [ \left(\frac{\sigma}{r}\right)^{12} - \left(\frac{\sigma}{r}\right)^6 \right ]
\end{equation*}
in the isobaric-isothermal ($NPT$) ensemble. 
Here $\sigma$ and $\epsilon$ set the length and energy scales, respectively, and $r$ is the distance between a pair of particles.
Molecular dynamics simulations with a Lennard-Jones pair potential were originally used to model liquid Argon,\cite{rahman_correlations_1964} but the simple form of the pair potential has also made it a principle testing ground for computational methods, including methods for nucleation and local structure detection.\cite{moroni_interplay_2005, lechner_accurate_2008, eslami_local_2017, bulutoglu_comprehensive_2023}
We truncate the pair potential at $r=2.5\sigma$ and apply isotropic tail corrections for the energy and pressure.  We integrate the equations of motion with a Nos\'e-Hoover thermostat and barostat, as  implemented in the LAMMPS molecular dynamics code.\cite{plimpton_fast_1995} We use a timestep $\Delta t = 0.004 \sqrt{m \sigma^2 / \epsilon}$, where $m$ is the particle mass, and relaxation constants of $500$ and $100$ timesteps for the barostat and thermostat, respectively.
Unless stated otherwise, the simulation box is cubic and the barostat is isotropic.

\subsection{Brute-force nucleation simulations}
We first perform brute-force simulations of nucleation at three different state points. These state points correspond to approximately $30\%$ supercooling at three different pressures: $P\sigma^3/\epsilon=0, ~5.68$, and $50$. The exact temperatures are $k_BT/\epsilon = 0.50,~ 0.79,$ and $2.30$, respectively. Here $P$ denotes the pressure, $T$ the temperature, and $k_B$ Boltzmann's constant. For these simulations, we start with a fluid phase and wait for the system to spontaneously crystallize. We perform $16$ independent brute-force nucleation simulations for each state point.


\subsection{Transition path sampling}
Similar to Ref.\ \citenum{ten_wolde_numerical_1995}, we focus on  temperature $k_BT/\epsilon=0.92$ and  pressure $P\sigma^3 / \epsilon = 5.68$. As the melting temperature at this pressure is approximately $k_BT_m/\epsilon=1.11$, \cite{van_der_hoef_free_2000} these conditions correspond to a supercooling of around $20\%$.

At this supercooling, there is a significant barrier for nucleation, meaning that it would take a very large amount of computer time for nucleation to occur spontaneously. To address this issue, enhanced sampling techniques such as umbrella sampling,\cite{auer_prediction_2001,torrie_nonphysical_1977} metadynamics,\cite{laio_escaping_2002,trudu_freezing_2006, eslami_local_2017} forward-flux sampling,\cite{allen_sampling_2005}  and transition path sampling,\cite{bolhuis_transition_2002,peters_obtaining_2006,lechner_role_2011,beckham_optimizing_2011,jungblut_crystallization_2011,moroni_interplay_2005} have been introduced.
All these techniques require an order parameter to monitor, or even drive, the progress of a nucleation trajectory. Previous work has used, for example, the number of particles in the nucleus, or the global Steinhardt bond orientational order $Q_6$ of the system,\cite{ten_wolde_numerical_1995} as the order parameter. In the case of metadynamics, the choice of order parameter directly 
influences the dynamics to favor sampling away from already visited configurations, whereas umbrella sampling restrains the sampling toward a small order parameter range.
In this work, we focus on the influence of the local structure detection algorithms on the \emph{analysis} of crystal nuclei, rather than on the potential influence of an order parameter on the \emph{simulation} of crystal nucleation. To minimize the 
impact of the choice of order parameter on the nucleation mechanism, we use transition path sampling.


 \begin{figure*}[t]
     \centering
     \includegraphics[width=\linewidth]{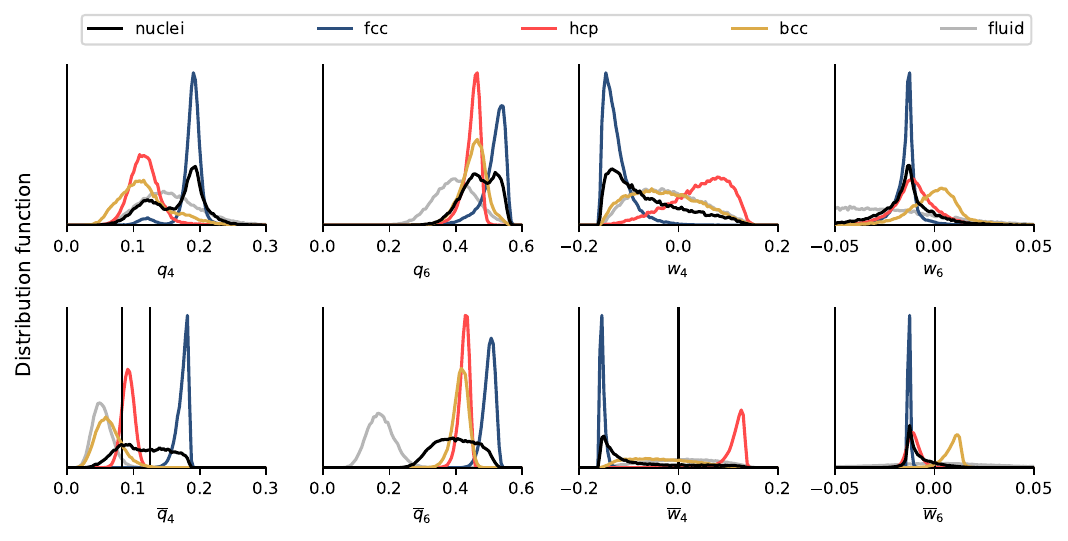}
     \caption{Distributions of the non-averaged bond order parameters  $q_4, q_6, w_4, w_{6}$ (top row) and their locally averaged counterparts $\q_4, \q_6, \w_4, \w_{6}$ (bottom row).~\cite{lechner_accurate_2008} The lines represent the distributions of bulk phases at $P\sigma^3/\epsilon=5.68$ and $k_BT/\epsilon = 0.92$, while the black lines correspond to particles in critical crystal nuclei under the same conditions. The vertical black lines in the bottom row indicate the thresholds used to distinguish between fcc, hcp, and bcc phases in the two polymorph classification schemes based on locally averaged bond order parameters.
     }
     \label{fig:bops-distributions}
 \end{figure*}

Transition path sampling (TPS) is designed to efficiently sample unbiased trajectories of rare events, such as nucleation. 
To initialize the procedure, we first require an initial nucleation trajectory, which we obtain from one of the brute-force nucleation simulations at $k_BT/\epsilon=0.79$. 
From this initial trajectory, TPS generates an unbiased ensemble of nucleation trajectories using `shooting moves'. These shooting moves generate a modified trajectory from a previous trajectory,
thereby performing a random walk in trajectory space in a Monte Carlo fashion,
and equilibrating the ensemble of trajectories. 
In \Cref{sec:tps-details}, we explain the details of our TPS simulations. The result of our TPS simulations is an unbiased ensemble of $50$ decorrelated nucleation trajectories.
Furthermore, we have performed a committor analysis to identify an ensemble of $50$ critical nuclei.

Using TPS, we have access to $50$ distinct critical nuclei and $50$ unique nucleation trajectories at a temperature of  $k_BT/\epsilon=0.92$ and a  pressure of $P\sigma^3/\epsilon = 5.68$. 

\section{Local structure detection methods} 
\label{sec:methods}
To analyze the local structure in all these nucleation trajectories, we employ a variety of local structure detection methods. In \Cref{sec:solid-like}, we explain how we identify crystal nuclei. Subsequently, we explain the seven different ways used to identify the crystal polymorphs face-centered cubic (fcc), hexagonal close-packed (hcp), and body-centered cubic (bcc) in these crystal nuclei. With the exception of polyhedral template matching,\cite{larsen_robust_2016} these methods are based on local order parameters derived from spherical harmonics expansions $q_{lm}(i)$ of the nearest neighbor density of each  particle $i$.\cite{steinhardt_bond-orientational_1983,ten_wolde_numerical_1995,lechner_accurate_2008}
Unless stated otherwise, nearest neighbors are identified using  the solid-angle based nearest neighbor algorithm.\cite{van_meel_parameter-free_2012} 
From the spherical harmonics expansion $q_{lm}(i)$, we compute the inner products $d_6 (i,j) = q_{6m}(i) \cdot q_{6m}(j)$ as introduced in Ref.\ \citenum{ten_wolde_numerical_1995}, along with the non-averaged bond order parameters $q_l(i)$ and $w_l(i)$, and the locally averaged bond order parameters $\q_l(i)$ and $\w_l(i)$ as introduced in Ref.\ \citenum{lechner_accurate_2008}. Please see \Cref{sec:bops} for the precise definition of these local order parameters.


\subsection{Identification of crystal nuclei}
\label{sec:solid-like}
Following Ref. \citenum{ten_wolde_numerical_1995}, we use the inner products $d_6(i,j)$ to identify crystal nuclei. A pair of particles $(i,j)$ has a `solid-like bond' if $d_6(i,j) > 0.7$. Subsequently, a particle is classified as `solid-like' if it has at least six of such solid-like bonds, and solid-like particles are considered to belong to the same solid-like cluster if they have a solid-like bond. 
For each snapshot in our simulations, we identify the largest solid-like cluster as the crystal nucleus. Only particles within this  crystal nucleus are further analyzed using a polymorph classification scheme, while all other particles are simply  labeled as `fluid-like'.

\subsection{Polymorph classification with non-averaged bond order parameters}
We start with the polymorph classification scheme used by \citeauthor{ten_wolde_numerical_1995}.\cite{ten_wolde_numerical_1995,ten_wolde_numerical_1996}
This scheme is based on the distributions of the non-averaged bond order parameters $(q_4,q_6,w_6)$.
Ref.\ \citenum{ten_wolde_numerical_1995} calculated histograms of $(q_4,q_6,w_6)$ and concatenated these histograms to form a `characteristic vector' $v$.  To classify a collection of particles, the characteric vector $v$ is calculated and 
projected onto the characteristic vectors of the bulk phases. To be specific, $v$ is decomposed using a linear least-squares algorithm into  contributions from the fcc, bcc, and fluid phases by minimizing the distance 
\begin{equation}
    \label{eq:hist-excl-hcp}
     \Delta^2 = \left [ v - (f_{\mathrm{fcc}} v_{\mathrm{fcc}} + f_{\mathrm{bcc}} v_{\mathrm{bcc}} + f_{\mathrm{fl}} v_{\mathrm{fl}}) \right ]^2.
\end{equation}
Here, $f_{\mathrm{fcc}}$, $f_{\mathrm{bcc}}$, $f_{\mathrm{fl}}$ represent the fractions of fcc, bcc, and fluid-like order, respectively. During the optimization, the fractions are constrained to lie between $0$ and $1$.

 We note that the presence of hcp-like order is not considered in Ref.\ \citenum{ten_wolde_numerical_1995}. To investigate its significance, we also examine a variant in which we project the characteristic vector onto the fcc, hcp, bcc, and fluid phases by minimizing the distance
 \begin{equation}
     \label{eq:hist-incl-hcp}
     \Delta'^2 = \left [ v - (f_{\mathrm{fcc}} v_{\mathrm{fcc}} + f_{\mathrm{hcp}} v_{\mathrm{hcp}} + f_{\mathrm{bcc}} v_{\mathrm{bcc}} + f_{\mathrm{fl}} v_{\mathrm{fl}}) \right ]^2.
\end{equation}
We will refer to the classification scheme based on \Cref{eq:hist-excl-hcp} as the `histogram scheme excluding hcp' and the one based on \Cref{eq:hist-incl-hcp} as the `histogram scheme including hcp'. 

To illustrate the histogram schemes, we present  the distributions of $q_4, q_6, w_4$, and $w_6$ in the top row of \Cref{fig:bops-distributions}. These distributions correspond to  the bulk phases at the simulation conditions $P\sigma^3/\epsilon=5.68$ and $k_BT/\epsilon = 0.92$. Although the distributions for the bulk phases are highly overlapping, there are distinct differences. For example, particles in the fcc phase generally exhibit  higher $q_4$ values compared  to  particles in the bcc phase.  The black line corresponds to the bond order distributions of  particles within  critical nuclei. To determine the relative contributions of  different phases in these nuclei, the black lines are fitted as a linear combination of the other distributions.

\subsection{Polymorph classification with locally averaged bond order parameters}
Next, we discuss two different polymorph classification schemes based on the locally averaged bond order parameters $\q_l$ and $\w_l$. We label these schemes according to the order parameters they are based on: the $(\w_4,\w_6)$-scheme and the $\q_4$-scheme.
In the $(\w_4,\w_6)$-scheme, a solid-like particle is classified as follows:
\begin{equation*}
\begin{cases}
     \textrm{fcc-like}  &\textrm{ if }  \w_6 < 0 \textrm{ and } \w_4 < 0,\\
     \textrm{hcp-like} &\textrm{ if } \w_6 < 0 \textrm{ and } \w_4 > 0, \\
     \textrm{bcc-like} &\textrm{ if }   \w_6 > 0.
\end{cases} 
\end{equation*}
In the $\q_4$-scheme, a solid-like particle is classified as follows:
\begin{equation*}
\begin{cases}
     \textrm{fcc-like}  &\textrm{ if }  \q_4 > 0.125, \\
     \textrm{hcp-like} &\textrm{ if }  0.083 < \q_4 < 0.125,  \\
     \textrm{bcc-like} &\textrm{ if }   \q_4 < 0.083. 
\end{cases} 
\end{equation*}
We have chosen these thresholds to identify  the crystal polymorphs  with the fewest mislabeled particles.\cite{espinosa_seeding_2016}

To illustrate these two schemes, we show the distributions of $\q_4, \q_6, \w_4$, and $\w_6$ in the bottom row of \Cref{fig:bops-distributions}. Again, these distributions correspond to the bulk phases and the critical nuclei.
The thresholds used above to distinguish the different phases are plotted as black vertical lines.
As Ref.\ \citenum{lechner_accurate_2008} demonstrated, the distributions
of the locally averaged bond order parameters $(\q_l,\w_l)$ are well-separated.
For example, the fluid phase is well-separated from the crystal phases with respect to $\q_6$, and the fcc and hcp crystal phase are well-separated based on  $\w_4$ and $\q_4$. 
Although $\q_4$ is commonly used to distinguish between fcc, hcp and bcc phases, we note that the distributions for hcp and bcc show significant overlap.

\begin{figure}[hbtp]
    \subfloat{
    \includegraphics[width=\linewidth]{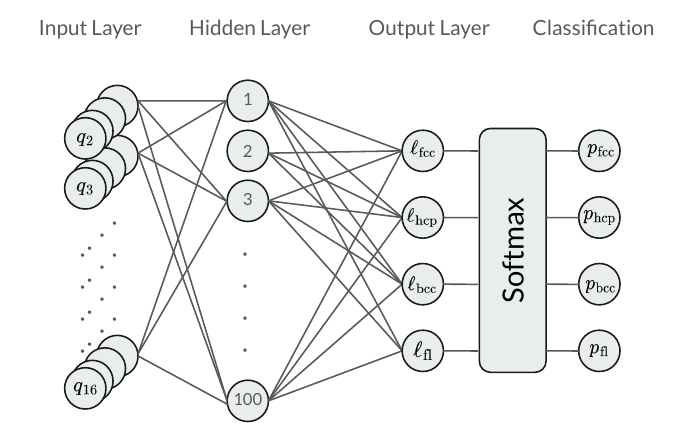}
    }
    \\
    \subfloat{
    \includegraphics[width=\linewidth]{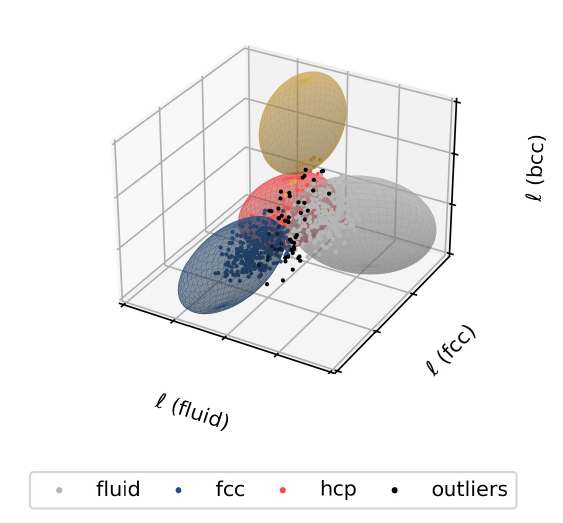}
    }
    \caption{Machine-learned polymorph classification with outlier detection. For each particle, $195$ different non-averaged bond order parameters $q_l$ are computed. A neural network transforms these $q_l$ into non-normalized log-likehoods $\ell$ for each bulk phase. In this four-dimensional latent space, multi-dimensional Gaussian distributions, here represented by ellipsoids, are fitted to the bulk phase distributions. If a particle environment maps to the interior of an ellipsoid, it is classified as belonging to the corresponding phase, otherwise, it is considered an outlier.
    }
    \label{fig:novelty-detection}
\end{figure}



\subsection{Machine-learned polymorph classification with outlier detection}
In recent years, there has been a rise in the use of machine learning for local structure detection.\cite{geiger_neural_2013,spellings_machine_2018,boattini_neural-network-based_2018,sdefever_generalized_2019,adorf_analysis_2020,coli_artificial_2021,leoni_nonclassical_2021,chung_data-centric_2022,becker_unsupervised_2022,terao_anomaly_2023}
Specifically, several approaches have been proposed for crystal polymorph classification. 
While we do not claim that the schemes used here are necessarily the best or the only methods for polymorph classification using machine learning, there are several  general aspects  we believe are important to consider when applying polymorph classification to nucleation.
First,  to compare with  other polymorph classification schemes, we seek  a machine-learning method that classifies particles as fluid, fcc, hcp, or bcc.
Second, given the uncertain nature of interfacial particles,  as observed in ice nucleation studies,\cite{espinosa_interfacial_2016,prestipino_barrier_2018,leoni_nonclassical_2021} the method should be able to  recognize when the local structure does not correspond to any of these phases.
Third, to avoid coarse-graining across solid-fluid or solid-solid interfaces, the method should rely solely on the nearest neighbors of a particle.
For these reasons, we use the following machine-learned polymorph classification schemes, which are largely  inspired by Ref.\ \citenum{chung_data-centric_2022}.

To capture the local structure around a particle $i$, we use the non-averaged bond order parameters $q_l(i)$, as they rely solely  on  nearest-neighbor information.
However, as we have seen in \Cref{fig:bops-distributions},  these distributions exhibit more overlap compared to their averaged counterparts.
Therefore, we employ a neural network to extract more informative features from the non-averaged bond order parameters.\cite{coli_artificial_2021} 
To be more precise, the neural network performs a non-linear combination and dimensionality reduction of the input features into a lower-dimensional space, where the bulk phases are classified using a Softmax function.
The neural network architecture is visualized in \Cref{fig:novelty-detection}.
The input to the neural network is a $195$-dimensional vector consisting of non-averaged bond order parameters $q_l(i)$ calculated using contributions from varying numbers of nearest neighbors.
To be more specific, for each number $k=2, \dots, 14$, we identify the $k$ nearest neighbors to particle $i$ and compute the spherical harmonics expansions $q_{lm}(i)$.
Then, for each number of nearest neighbors $k$, we compute the bond order parameters $q_l$ where we vary $l$ from $2$ to $16$. In this way, we compute $13$ sets of bond order parameters $q_2, \dots, q_{16}$, giving a total number of $195 = 13\times 15$ input features. We train the neural network to classify  bulk fluid and crystal phases based on these input features. The neural network consists of $195$ input nodes, one hidden layer of $100$ nodes, and four output nodes corresponding to the four different phases. We also considered neural network architectures with more hidden layers and more hidden nodes, but found that these changes did not significantly improve the classification accuracy. Likewise, we considered including the third-order invariants $w_l$ to the input features, but this also did not significantly improve the classification accuracy. In \Cref{sec:sensitivity}, we perform a sensitivity analysis to assess which of the $195$ input features are  most important for the classification problem.

For the training data, we consider two approaches. In the first approach, following Ref.\ \citenum{chung_data-centric_2022}, we generate the training data by adding Gaussian noise to ideal crystal lattices, specifically for fcc, hcp, and bcc. In other words, the training data is derived from Einstein crystals.
The spring constants of these Einstein crystals are set to match those  measured for a thermally equilibrated fcc crystal at our simulation conditions. 
The average magnitude of the Gaussian noise is approximately $10\%$ of the nearest neighbor distance in each case.
Additionally, we add an equilibrated bulk fluid phase in the training data to enhance  the network's capability to  distinguish fluid-like from crystalline local structures.
In the second approach,  the training data is simply obtained from thermally equilibrated fcc, hcp, bcc and fluid phases at our simulation conditions.
We refer to the machine-learned (ML) classifiers based on these two training sets as the ML1 and ML2 schemes, respectively. To be clear,  the ML1 scheme uses ideal crystal lattices with noise, while the ML2 scheme uses thermally equilibrated crystal phases.


As mentioned above, we require a  method to detect when the local structure does not correspond to any of these phases.
In machine learning, identifying deviations from a set of reference data is referred to as novelty detection or outlier detection.
To describe the outlier detection algorithm that we use, we first note that the output of the neural network is a non-normalized log-likelihood $\ell$ for each phase. For example, the probability $p_{\mathrm{fcc}} $ that a particle belongs to the fcc phase is computed as 
\begin{equation*}
    p_{\mathrm{fcc}} = \frac{e^{\ell_{\mathrm{fcc}}}}{ e^{\ell_{\mathrm{fcc}}} + e^{\ell_{\mathrm{hcp}}} + e^{\ell_{\mathrm{bcc}}} + e^{\ell_{\mathrm{fl}}}}.
\end{equation*}
The set of non-normalized log-likelihoods $\ell$ provides a convenient `latent space' in which we can apply the outlier detection scheme proposed in Ref. \citenum{lee_simple_2018}.
Each bulk phase corresponds to a specific distribution within this four-dimensional latent space.
By inspecting the survival functions of the so-called `Mahalanobis' distance, we find that these distributions are well-described by multi-dimensional Gaussian distributions.
In \Cref{fig:novelty-detection}, we visualize three dimensions of the four-dimensional latent space. The Gaussian distributions corresponding to the bulk phases are represented by ellipsoids. The  surface of each ellipsoid represents, 
within the Gaussian approximation,  the smallest surface that encloses $95\%$ of the distribution.
Consequently, $95\%$ of the particles in a bulk fcc phase will be mapped within the interior of the fcc ellipsoid. 
We have chosen this threshold of $95\%$ conservatively to ensure high confidence in positive classifications. Note that we use the outlier detection scheme on the latent space of log-likelihoods $\ell$ rather than using a threshold on the classification probabilities $p_{\mathrm{fcc}}, p_{\mathrm{hcp}}, \dots$. This approach is chosen because the classification probabilities output by a neural network can be misleadingly  high even for outliers.\cite{lee_simple_2018} For example, when classifying a simple cubic crystal structure using the ML1 neural network, the output probability was $p>0.99$ for the hexagonal close-packed (hcp) class, even though the simple cubic structure is clearly  different from the hexagonal close-packed structure. In contrast, our outlier detection method correctly identifies  the simple cubic crystal structure as an outlier.

In summary, our machine-learned polymorph classification schemes with outlier detection, ML1 and ML2, operate similarly, with the only difference between them being  the training data. For each particle, $195$  non-averaged different bond order parameters $q_l$ are calculated. These $q_l$ values are then mapped by the neural network to a four-dimensional latent space where the bulk phases are separated. If a particle in this space falls within the ellipsoid corresponding to a specific phase, we classify this particle as belonging to that phase. If this particle lies outside  all ellipsoids, we classify this particle as an outlier. In \Cref{fig:novelty-detection}, we show an example where the ML1 scheme is applied to a nucleus from our TPS simulations. Each small dot represents  a single particle within the nucleus, and the dots are colored according to their classification, with black dots indicating  outliers.


\subsection{Polyhedral Template Matching}
\begin{figure}
    \centering
    \includegraphics[width=\linewidth]{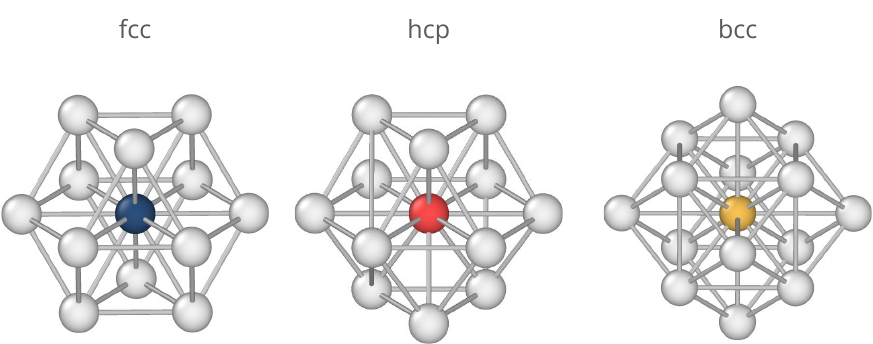}
    \caption{Templates used for the identification of local fcc, hcp, and bcc crystal structures in the polyhedral template matching (PTM) algorithm.}
    \label{fig:ptm-templates}
\end{figure}

Finally, we also employ a polymorph classification method known as  polyhedral template matching (PTM).\cite{larsen_robust_2016}
The PTM algorithm identifies  local crystal structure by comparing the real-space positions of a central particle and its nearest neighbors to a predefined reference template. The reference templates used  for fcc, hcp, and bcc in PTM are visualized in \Cref{fig:ptm-templates}.
For fcc and hcp, the template consists of the central particle and its twelve nearest neighbors, while for bcc, the template contains the central particle and its fourteen nearest neighbors. 
The output of the PTM algorithm is a root-mean-square-deviation (RMSD) score for each particle relative to each crystal template.
The particle is classified according to the crystal phase with the lowest RMSD score, indicating  the best-matching crystal phase. 
In order to increase confidence in the classification, Ref.\ \citenum{larsen_robust_2016} recommends applying a maximum value of the RMSD. 
We use a maximum RMSD of $0.12$; thus, if a particle's RMSD exceeds this value, it is classified as fluid-like. 

\begin{figure*}[t]
     \centering
     \includegraphics[width=0.97\linewidth]{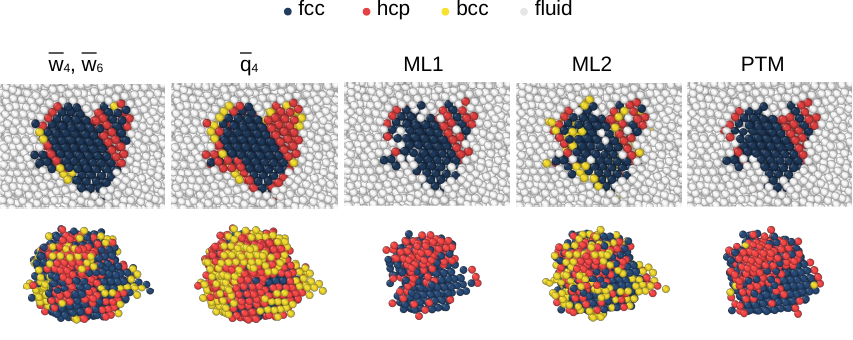}
     \caption{Two crystal nuclei classified using five different polymorph classification schemes. The labels correspond to the polymorph classification methods described in \Cref{sec:methods}. The top row shows a cut-through image of a nucleus of around $600$ particles, while the bottom row shows the external surface of a post-critical nucleus of around $1100$ particles.}
     \label{fig:nuclei-snaps}
 \end{figure*}
 
\section{Results}
We apply the seven different polymorph classification schemes as introduced in \Cref{sec:methods}, to the crystal nuclei  sampled with transition path sampling. To summarize, these schemes include two variants of the histogram method used by \citet{ten_wolde_numerical_1995},  one including hcp and one excluding hcp; 
two schemes based on Lechner-Dellago's locally averaged bond order parameters\cite{lechner_accurate_2008}: the $(\w_4,\w_6)$- and $\q_4$-schemes;
two machine learning methods with outlier detection inspired by Ref.\ \citenum{chung_data-centric_2022},  labeled as ML1 and ML2;
and finally the polyhedral template matching (PTM) algorithm.\cite{larsen_robust_2016}

To illustrate these schemes, we start by applying them to two representative nuclei from the TPS simulations. In \Cref{fig:nuclei-snaps}, the color of the particles indicates their local structure: fcc-like particles are dark blue, hcp-like particles are red, bcc-like particles are yellow, and fluid-like particles are light gray. 
The top row shows a cut-through image of a nucleus around the critical size of $600$ particles, while the bottom row shows the external surface of a post-critical nucleus of around $1100$ particles. For this post-critical nucleus, fluid-like particles are not shown.
Note that the histogram-based schemes inherently apply to  collections  of particles, rather than single particles, which is  why they are not included  in \Cref{fig:nuclei-snaps}. 
All classification schemes agree that the core of the nucleus is predominantly fcc-like. 
However, they differ in how they classify the local structures at the interface between the crystal nucleus and the surrounding fluid. 
This disagreement is especially visible in the bottom row, where the fluid-crystal interface of the post-critical nucleus looks completely different for each different classification scheme.
According to the $(\w_4, \w_6)$-scheme, the interface is mainly fcc-like with smaller amounts of hcp and bcc. In contrast the $\q_4$ and ML2 schemes identify a higher portion of hcp- and bcc-like particles at the interface. 
More remarkably, both  the PTM and ML1 schemes identify almost no bcc-like particles within these nuclei.
Interestingly, the methods not only disagree near the solid-fluid interface but also near  internal solid-solid interfaces. For instance, the top right of the smaller nucleus (top row of \Cref{fig:nuclei-snaps}) is predominantly classified as hcp-like according to the $\q_4$-scheme. In contrast, all other methods agree that this region contains two hcp-like layers in between fcc-like layers. This example illustrates how  classification schemes can disagree on the identification of local structures, particularly near solid-solid and solid-fluid interfaces.

\begin{figure*}
 \centering
\includegraphics[height=205mm]{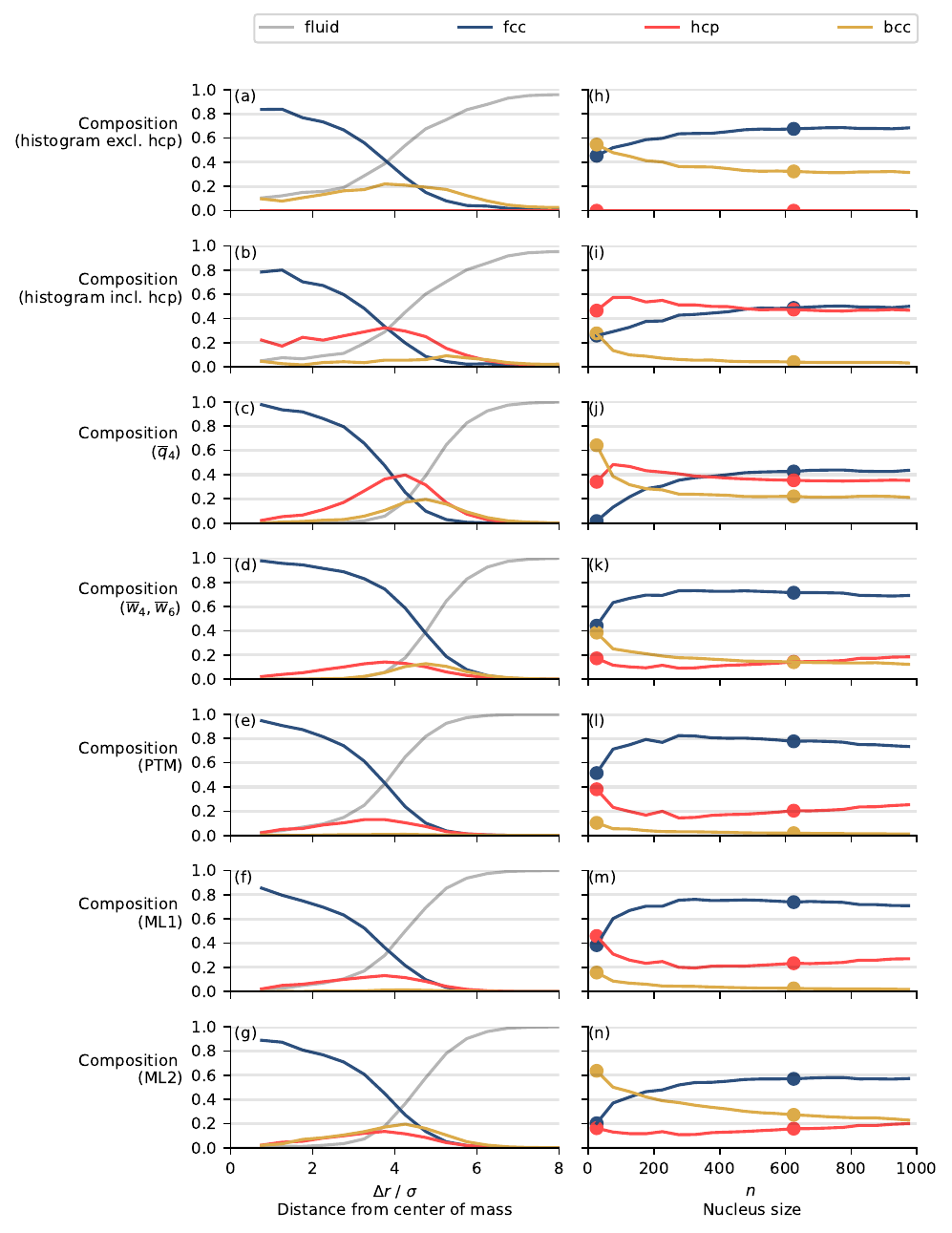}
 \caption{Polymorph composition of nuclei during crystallization from a Lennard-Jones fluid at pressure $P\sigma^3/\epsilon=5.68$ and temperature $k_BT/\epsilon=0.92$ as classified by seven different local structure detection methods. (a-g) Composition of critical nuclei as a function of the distance $\Delta r/\sigma$ from the center of mass of the nuclei. (h-n) Composition of nuclei as a function of the nucleus size, i.e.\ the number of particles in the nuclei. The dots in (h-n) highlight two selected nucleus sizes: $n\approx 25$ and $n\approx 625$. The latter corresponds approximately to the critical nucleus size. 
 Please see \Cref{sec:methods} for an explanation of the local structure detection methods.
 }
\label{fig:compositions}
\end{figure*}

\subsection{Radial dependence of critical nucleus composition}
To provide a more quantitative analysis, we examine the local structure of the ensemble of critical nuclei.
For each particle in the system, we determine its local structure using the various classification schemes and measure its distance from the center of mass of the critical nucleus. 
In \Cref{fig:compositions}(a-g), we plot the fraction of particles  classified as fcc-, hcp-, bcc-, and fluid-like, as a function of their distance from the center of mass of the critical nucleus.

We observe the same trends as in \Cref{fig:nuclei-snaps}: the core of the nuclei is predominantly fcc-like, while the classification of the interface varies significantly among the different schemes. 
In line with the findings of Ref.\ \citenum{ten_wolde_numerical_1995}, the histogram scheme excluding hcp shows  a significant presence of bcc  on the surface. For sufficiently large  distances $\Delta r/\sigma$ from the center of mass, the bcc fraction exceeds even  the fcc fraction. In contrast, the histogram scheme including hcp suggests that hcp ordering predominates the nucleus surface, with the bcc fraction  significantly reduced  but  still larger than the fcc fraction for larger  $\Delta r/\sigma$. The $\q_4$-scheme qualitatively agrees with the histogram scheme including hcp.
On the other hand, the $(\w_4,\w_6)$, PTM and ML1 schemes agree that the fcc structure still dominates the surface, but they differ regarding  the bcc fraction: PTM and ML1 indicate that bcc is almost completely absent, while the $(\w_4,\w_6)$-scheme still detects a significant bcc fraction near the nucleus surface. ML2 shows a dominant bcc and a significant hcp presence on the surface.

In summary, we  conclude that the solid-fluid interface of critical nuclei is bcc-dominated according to the histogram scheme excluding hcp and ML2, hcp-dominated according to the  histogram scheme including hcp and $\q_4$, and fcc-dominated according to $(\w_4,\w_6)$, PTM and ML1. The only consistent finding all methods agree on is that the core is fcc-dominated.



\subsection{Size dependence of nucleus composition}
These findings also have implications for the size dependence of the nucleus composition. We divided the crystal nuclei from our TPS simulations into $20$ different groups based on their nucleus size, with each group corresponding to nucleus sizes ranging from $0-50$, $50-100$, ..., $950-1000$. For each group, we classified the local structure according to the seven polymorph classification schemes and averaged the fractions of fcc-, hcp-, bcc-, and fluid-like order in each group.
In \Cref{fig:compositions}(h-n), we plot the resulting averaged composition as a function of  nucleus size. To facilitate comparison, we have  normalized the composition so that the fcc, hcp, and bcc fractions sum to one. Furthermore, we have highlighted two nucleus size groups with dots: the smallest nuclei with sizes $0-50$ and nuclei around the critical size with sizes $600 - 650$. We first discuss the composition of the smallest nuclei, and subsequently,  the composition of the critical nuclei.

For the smallest nuclei, the histogram scheme excluding hcp  suggests that bcc dominates, consistent with the findings of  Ref.\ \citenum{ten_wolde_numerical_1995}. The $\q_4$ and ML2 schemes also indicate that bcc dominates, while the histogram scheme including hcp and ML1 indicate that hcp dominates. In contrast, the $(\w_4, \w_6)$ and PTM schemes agree that fcc  dominates even in the smallest nuclei. The disagreement between these different schemes becomes  especially clear when ranking  the fractions of fcc, hcp and bcc. 
For the smallest nuclei, nearly  every possible ordering of the six combinations,  (1)  fcc $>$ hcp $>$ bcc, (2) fcc $>$ bcc $>$ hcp, (3) hcp $>$ fcc $>$ bcc, etc., is obtained as almost every polymorph classification scheme yields a different ranking. 

For the critical nuclei, all classification schemes agree that fcc dominates. The hcp fraction is almost equal to the fcc fraction according to the histogram scheme including hcp  and the $\q_4$-scheme, but significantly smaller according to the histogram scheme excluding hcp, the $(\w_4,\w_6)$, PTM, and ML schemes. The bcc fraction is 
approximately $30\%$ according to the histogram scheme excluding hcp and ML2, around $20\%$ using the $(\w_4,\w_6)$ and $\q_4$ schemes, and less  than $5\%$ according to the histogram scheme including hcp  ($4\%$), ML1 ($3\%$), and PTM ($2\%$). Although the bcc fraction is low according to PTM and ML1, it is still non-zero.


In summary, the size-dependence of the nucleus composition varies significantly depending on the polymorph classification scheme.
For the smallest precritical nuclei, nearly  every scheme disagrees on the relative importance of fcc, hcp and bcc structures. While most schemes agree that fcc order dominates in  the critical nuclei, they diverge on the relative importance of hcp and bcc.

\subsection{Pressure and temperature dependence of nucleus composition}
So far, we have observed that the composition of crystal nuclei can depend sensitively on the choice of polymorph classification scheme. In this section, we explore a completely different factor that can influence the composition of crystal nuclei: the state point. As shown in Ref.\ \citenum{desgranges_controlling_2007},  the composition of post-critical Lennard-Jones nuclei varies widely with  pressure and temperature. To be specific, Ref.\ \citenum{desgranges_controlling_2007} found that the hcp fraction increases with supercooling, while the bcc fraction increases  with pressure. At sufficiently  high pressures, the post-critical nuclei were found to be almost entirely bcc-like.

\begin{figure}
    \centering
    \includegraphics[width=\linewidth]{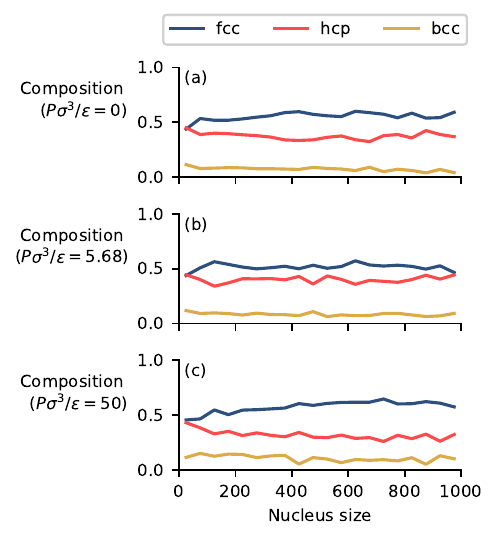}
    \caption{Pressure and size dependence of the polymorph composition of Lennard-Jones crystal nuclei. For each pressure, we performed brute-force simulations to obtain nucleation trajectories at approximately $30\%$ supercooling.
    The local crystal structure in this analysis is determined using polyhedral template matching.\cite{larsen_robust_2016}
    }
    \label{fig:spontaneous}
\end{figure}

To investigate this, we analyze the brute-force simulations that we performed at $30\%$ supercooling at three different pressures.
The three pressures are $P\sigma^3/\epsilon=0,~5.68$ and $50$.
As before, we divide the crystal nuclei from the brute-force simulations into $20$  groups based on their nucleus size, ranging from $0-50$, $50-100$, up to $950-1000$.
We classify the local structure using the polyhedral template matching (PTM) algorithm.\cite{larsen_robust_2016} 
We chose PTM because we consider  it to be the most reliable method.
Furthermore, together with the $(\w_4,\w_6)$-scheme, PTM is the only scheme that can be automatically transferred  to different state points.
For each nucleus size group, we compute the average fraction of fcc-, hcp-, bcc-, and fluid-like order. In \Cref{fig:spontaneous}, we plot the average composition as a function of the nucleus size. Again, we normalize the composition so that the fcc, hcp, and bcc fractions sum to one.

The composition of crystal nuclei appears to be largely  independent of both nucleus size and pressure. Fcc consistently  dominates, followed by hcp, and then bcc. The smallest nuclei exhibit a nearly equal fraction of fcc and hcp. Quantitatively, the fcc fraction fluctuates around $55\%$, the hcp fraction around $36\%$, and the bcc fraction around $9\%$. The average bcc fraction increases slightly with pressure: for $P\sigma^3/\epsilon=0$  it fluctuates around $7\%$, for $P\sigma^3/\epsilon=5.68$ around $8\%$, and for $P\sigma^3/\epsilon=50$ around $11\%$. For higher pressures, the hcp fraction decreases slightly to around $32\%$.

We were unable to reproduce the completely bcc-like crystallites observed in Ref.\ \citenum{desgranges_controlling_2007} at high pressures.
This discrepancy could stem from differences in the  polymorph classification schemes used. Ref.\ \citenum{desgranges_controlling_2007} used thresholds for the non-averaged bond-order parameters $q_4$ and $w_4$ to distinguish fcc, hcp and bcc.\cite{desgranges_molecular_2006} Due to the significant overlap in these order parameter distributions, as shown in \Cref{fig:bops-distributions}, we were unable to reproduce their classification scheme effectively.
We also note that the formation of entirely bcc-like crystallites seems improbable in the light of the mechanical instability of the  bcc phase of Lennard-Jones particles.\cite{eshet_new_2008, schwerdtfeger_cuboidal_2022} 

To discuss the temperature dependence, we compare \Cref{fig:spontaneous}(b) with \Cref{fig:compositions}(l). In this way, we are able to compare $30\%$ to $20\%$ supercooling at $P\sigma^3/\epsilon=5.68$ using the same polymorph classification scheme (PTM). We observe that the hcp and bcc fractions increase with supercooling, while the fcc fraction decreases with supercooling. Quantitatively, at $30\%$ supercooling, the fcc, hcp and bcc fractions  are approximately $55\%$, $36\%$ and $9\%$, respectively, as mentioned above. In contrast, at  $20\%$ supercooling, these fractions change to around $78\%$, $20\%$, and $2\%$. This  finding, particularly the increase in the hcp fraction  with supercooling, is in agreement with Ref. \citenum{desgranges_controlling_2007}.

Given the sensitivity of  composition to the polymorph classification scheme, all these fractions should be taken with a grain of salt.
For instance, using the $(\w_4,\w_6)$ scheme on the same brute-force nucleation trajectories reveals lower hcp and higher bcc fractions compared to the PTM scheme.
Nevertheless, both the PTM and $(\w_4,\w_6)$ scheme agree on the following qualitative conclusions.
Although the composition of crystal nuclei does not depend sensitively on  nucleus size or pressure, the hcp fraction decreases and the bcc fraction increases with pressure.
Furthermore, the hcp and bcc fractions appear to increase with supercooling, though they  remain smaller than the fcc fraction.

\section{Discussion}
\label{sec:discussion}

Why do   different polymorph detection schemes yield such varying results for the surface of crystal nuclei?
First, particles near a solid-fluid or solid-solid interface have a local  structure that  differs from that of  bulk fluid or bulk crystal phases. As a result, it is not immediately clear  whether order parameters that work well for distinguishing bulk phases will also be reliable  for classifying these interfacial particles.
It is possible that interfacial particles should not be classified with the same criteria as bulk phases and might  be better considered as distinct from the bulk altogether.
We will discuss this point in more detail below.
Additionally, we will explore how the choice of training data influences the analysis, including  which phases are considered and how  reference structures are obtained.

\subsection{Interfacial particles as outliers}
Since interfacial particles can differ significantly from bulk phases, it is useful to employ  an outlier detection method to identify when the local structure deviates from a bulk phase. 
The PTM and ML methods we examined both rely on the similarity of the positions of the nearest neighbors of a particle  to those in a reference crystal phase.
Essentially, these methods  compute a distance from the reference crystal phase and classify particles as outliers if this distance exceeds a certain  threshold.
Other methods based on templates such as common neighbor analysis,\cite{stukowski_structure_2012} the topological cluster classification,\cite{malins_identification_2013} and analysis of Voronoi polyhedra,\cite{swope_106-particle_1990} are even stricter: they require the local (topological) structure to match   the reference template exactly.
Note that \citet{swope_106-particle_1990}  also found a small fraction of bcc-like particles in Lennard-Jones nucleation using the Voronoi polyhedra analysis.
It is somewhat reassuring that our PTM and ML1 results align  with those in Ref.\ \citenum{swope_106-particle_1990} on this point. All three methods  use some kind of outlier detection method and rely solely on the nearest neighbors of a particle. They  all find a small fraction of bcc-like particles in Lennard-Jones nucleation.

\begin{figure*}[tbp]
    \includegraphics[width=\linewidth]{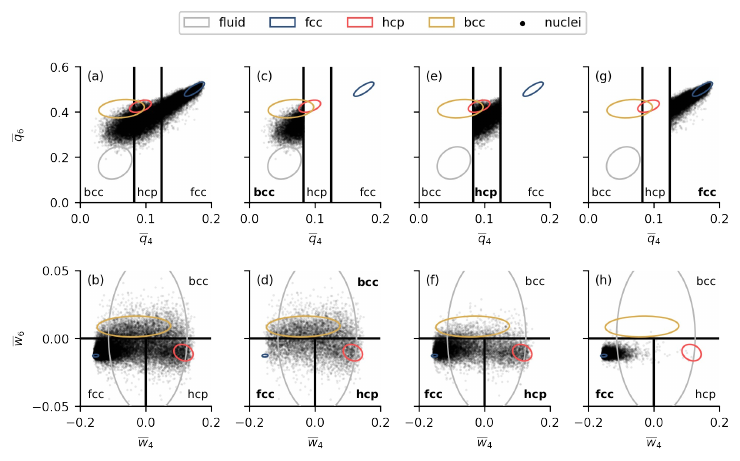}
    \caption{Two-dimensional scatter plot of the averaged bond order parameters $(\q_4, \q_6)$ and $(\w_4, \w_6)$ for particles in the critical nuclei from the transition path sampling simulations at $P\sigma^3/\epsilon=5.68$ and $k_BT/\epsilon=0.92$. The ellipses enclose $95\%$ of the distributions of thermally equilibrated bulk phases at the same conditions. Horizontal and vertical black lines indicate the thresholds used to classify fcc, hcp, and bcc polymorphs in the $\q_4$ and $(\w_4,\w_6)$-schemes. 
    The columns show different subsets of the solid-like particles in the critical nuclei: all particles in the critical nuclei (a,b), particles  classified as bcc-like (c,d), hcp-like (e,f) and fcc-like (g,h)
according to the $\q_4$ 
-scheme. }
    \label{fig:2d-scatter}
\end{figure*}


The notion that particles near interfaces and within small crystal nuclei exhibit a distinct local structure is well-established in the literature.
This phenomenon, and related behaviors, are commonly referred to as pre-ordering,\cite{menon_role_2020} pre-structuring,\cite{lechner_role_2011} pre-cursors,\cite{russo_new_2014} or two-step nucleation.\cite{diaz_leines_maximum_2018}
The local structure of precursors or interfacial particles is frequently associated with a metastable bulk phase, such as a liquid,\cite{van_meel_two-step_2008} an amorphous solid,\cite{arjun_unbiased_2019} or a metastable crystal polymorph.\cite{russo_new_2014}
In contrast, several studies have also shown that the local structure of interfacial particles can differ from any bulk phase.
For example, it has been demonstrated that the interface possesses a distinct local topological structure compared to the bulk.\cite{menon_role_2020,becker_unsupervised_2022,gispen_crystal_2023} 
Additionally, interfacial particles exhibit values of $\q_4$ and $\q_6$ that differ from those of bulk phases.\cite{lechner_role_2011, diaz_leines_atomistic_2017, espinosa_interfacial_2016, prestipino_barrier_2018}


The latter effect is also evident in two-dimensional scatter plots of $(\q_4, \q_6)$ and $(\w_4, \w_6)$ shown in \Cref{fig:2d-scatter}(a,b), where the typical order parameter values for bulk phases are represented by  ellipses that contain $95\%$ of their distributions. The black dots represent the order parameter values of the solid-like particles in the critical nuclei. The black lines correspond to the thresholds used in the $(\w_4,\w_6)$- and $\q_4$-schemes for polymorph classification.
We observe that many particles exhibit order parameter values that lie outside  the bulk phase ellipses. Instead, these values often lie in an  intermediate range  between those of the fcc and fluid phase.



The observation that interfacial particles exhibit  $\q_4$ and $\q_6$ values distinct from bulk phases has prompted some authors \cite{lechner_role_2011,menon_role_2020} to adopt an outlier detection approach based on these order parameters.
In these studies, a particle is  classified as corresponding to a bulk phase only if its $\q_4$ and $\q_6$ values overlap directly  with the distribution of the respective bulk phase.
Roughly speaking, this `overlap criterion' requires that a black dot in \Cref{fig:2d-scatter}a would fall within  one of the ellipses, rather than in the intermediate white space.
It is reasonable to expect that this method likely results in a stricter and more reliable classification.
This overlap criterion resembles  the outlier detection algorithm employed in our machine learning method.
However, there are two key differences. 

The first difference is that we use non-averaged bond order parameters instead of locally averaged bond order parameters, as we believe that the overlap criterion may  be too strict when applied to locally averaged bond order parameters.
Locally averaged bond order parameters\cite{lechner_accurate_2008} characterize the local structure of a particle by incorporating both its  nearest neighbors  and the  neighbors of those nearest neighbors.
The computation of these order parameters effectively results in  a coarse-graining over a small surrounding region.
Consequently, applying a strict overlap criterion to locally averaged bond order parameters would require that the positions of all nearest neighbors and their  neighbors closely resemble  those in a bulk phase.
In a crystal nucleus, such a requirement would likely exclude  particles on the surface, and possibly the layer below the surface, from being classified as part of the bulk phase. 
Near solid-solid interfaces, such as in a random-hexagonal close packed (rhcp) stacking, such a requirement would also exclude many particles from being classified.
For example, in \Cref{fig:nuclei-snaps}, the top right of the nucleus would not be recognized as either fcc or hcp.
Indeed, as shown in \Cref{fig:2d-scatter}(a), nearly all particles fall outside  the ellipses.

The second difference is that we require overlap in a four-dimensional space rather than a two-dimensional space.
In one- or two-dimensional spaces, it is relatively easy to get overlap, and therefore relatively easy to identify metastable crystal phases on the surface of crystal nuclei.
As an extreme example, if we consider the one-dimensional $\q_4$-distributions in \Cref{fig:bops-distributions}, we see that the fluid phase  nearly completely overlaps with the bcc phase. Interfacial particles with values  between the fcc and fluid phases will inevitably  adopt intermediate $\q_4$ values, and consequently,  they must overlap with the hcp phase. This may explain  why hcp is frequently identified on the surface of crystal nuclei.\cite{kawasaki_formation_2010,lechner_role_2011,tan_visualizing_2014} A similar effect occurs with $\q_6$: interfacial particles with intermediate $\q_6$-values will inevitably overlap with the bcc or hcp phases. 

The two-dimensional scatterplots of $(\q_4, \q_6)$ and $(\w_4, \w_6)$ in \Cref{fig:2d-scatter}(c-h) also illustrate how different projections can lead to very different classifications. In these figures, we alternately plot the solid-like particles in the critical nuclei that are classified as bcc-like (c,d), hcp-like (e,f) and fcc-like (g,h) according to the $\q_4$-scheme. 
Particles that are classified as bcc-like by the $\q_4$-scheme show a wide variation of $\w_4$ and $\w_6$ values in \Cref{fig:2d-scatter}(d). The $(\w_4,\w_6)$-scheme would classify these particles as a distribution of fcc, hcp and bcc. They have almost no overlap with the bcc, hcp, or fcc ellipses in \Cref{fig:2d-scatter}(c), but have significant overlap with the bcc and hcp ellipses in \Cref{fig:2d-scatter}(d). Particles that are classified as hcp-like by the $\q_4$-scheme tend to have more negative $\w_6$ values  but still display a wide variation in $\w_4$. These particles are very far from the fcc ellipse in \Cref{fig:2d-scatter}(e), but  overlap with the fcc ellipse in \Cref{fig:2d-scatter}(f). Particles that are classified as fcc-like by the $\q_4$-scheme are 
usually also fcc-like according to the $(\w_4, \w_6)$-scheme.
These examples show that whether interfacial particles overlap with a bulk phase distribution can depend sensitively on the choice of order parameters, as was also shown for water in Ref.\ \citenum{espinosa_interfacial_2016}.

\subsection{Influence of training data}
So far, we have focused on how the choice of order parameters affects  the polymorph classification of interfacial particles. 
Another fundamentally different aspect of polymorph classification methods is the `training data' they rely on. By training data, we mean the following: which phases are considered in the analysis, and how are reference structures obtained?

Among the polymorph classification schemes we consider, the histogram, $(\w_4,\w_6)$, and $\q_4$ schemes are all based on thermally equilibrated bulk phases at the same simulation conditions as our transition path sampling simulations. However, the histogram scheme excluding hcp does not incorporate the hcp phase in its training data, while the histogram scheme including hcp does. Apart from this difference, both histogram schemes are exactly the same. We have seen how the inclusion or exclusion of hcp in the training data leads to significant differences in the classification of crystal nuclei. In particular, the bcc fraction decreases substantially, and the hcp fraction significantly increases when hcp is included in the histogram scheme.

The choice of training data also has a large impact on the results of our machine learning (ML) schemes.
The training data for our ML1 scheme is obtained by adding Gaussian noise to ideal fcc, hcp and bcc lattices.
The magnitude of the Gaussian noise is chosen such to resemble the thermal fluctuations at the relevant simulation conditions.
In contrast, the ML2 scheme directly uses training data from thermally equilibrated fluid, fcc, hcp and bcc phases under the same  simulation conditions.
The change in training data causes ML2 to identify a significantly larger bcc fraction compared to ML1.
Determining which of the two ML schemes is more reliable is open to interpretation. However, we consider the ML1 scheme as more reliable for the Lennard-Jones system, given the mechanical instability of the bcc phase.\cite{schwerdtfeger_cuboidal_2022} 
We find that the thermally equilibrated bcc phase exhibits much larger fluctuations around its ideal lattice positions compared to  the fcc and hcp phases.
In addition to its mechanical instability,\cite{schwerdtfeger_cuboidal_2022} these larger fluctuations in the bcc phase may also be related 
to the metastable $I\overline{4}3d$ phase.\cite{eshet_new_2008}
We refer readers to \Cref{sec:bcc-stability} for a more detailed discussion of the stability of bcc and the $I\overline{4}3d$ phases.
However, we note that  the bcc phase appears to transform into a heterogeneous mixture of fcc, hcp, and bcc structures.
As a result, the ML2 scheme may inadvertently incorporate locally fcc- and hcp-like structures into the bcc training data, potentially introducing  a bias toward bcc classification.

In summary, we have demonstrated  that the choice of training data can significantly influence  polymorph classification schemes, both for  histogram and ML schemes. 
In both cases, different training data lead to qualitatively different conclusions for the nucleation mechanism. 

\section{Conclusion}
In conclusion, we have shown that different local structure detection methods yield markedly different results for crystal nucleation from a Lennard-Jones fluid.
While all seven  methods we considered agree that the core of the critical nucleus is predominantly fcc-ordered, they 
vary in their assessment of  the surface composition of the critical nucleus, which can be   dominated by fcc, hcp or bcc, depending on the local structure detection method used.
Similarly, there is significant  disagreement among the methods regarding  the relative importance of fcc, hcp and bcc for the smallest precritical nuclei.
We attribute the discrepancies to the selection of reference crystal structures, the use of outlier detection, and the choice of local order parameters.

These results suggest that one should be very careful when characterizing the local structure near solid-solid interfaces,  solid-fluid interfaces, and  small crystal nuclei.
Particles near such interfaces often have  local  structures that deviate from those in  bulk fluid or bulk crystal phases.
Therefore, we advocate for incorporating outlier detection in local structure detection methods, as it provides  a stricter definition of crystallinity and enhances confidence in the classification results.


The two methods that we consider to be the most reliable are a machine learning method (ML1) with outlier detection and polyhedral template matching.\cite{larsen_robust_2016}
Both methods employ a strict definition of crystallinity and are the only ones that provide consistent results for the polymorph composition of Lennard-Jones crystal nuclei.
According to these more stringent  methods, there is virtually no bcc-like ordering on the surface of critical nuclei. The prominent role of metastable bcc ordering observed during Lennard-Jones crystal nucleation\cite{ten_wolde_numerical_1995,desgranges_controlling_2007} may be attributed to less rigorous local structure detection methods.




\section*{Supplementary material}
The supplementary material contains the code used to generate and analyze the results of this paper as well as the nucleation trajectories obtained using brute-force and transition path sampling simulations.

\begin{acknowledgments}
M.D., A.P.A.O. and W.G. acknowledge funding from the European Research Council (ERC) under the European Union’s
Horizon 2020 research and innovation programme (Grant
agreement No. ERC-2019-ADG 884902 SoftML).
We thank Peter Bolhuis for useful discussions, and Bryan Verhoef and Gabriele Coli for their contributions to an early version of this project.
\end{acknowledgments}

\section*{Data Availability Statement}
The data supporting the findings of this study are available within the article and its supplementary material.

\appendix
\section{Transition path sampling}
\label{sec:tps-details}
In this appendix, we provide  details of our transition path sampling (TPS) simulations  using the aimless shooting variant.\cite{peters_obtaining_2006} 
In each shooting move, a `shooting point', i.e.\ a time slice, is selected from the previous trajectory. The velocities of all  particles at this shooting point are completely resampled from a  Maxwell-Boltzmann distribution. From this shooting point with resampled velocities, the equations of motion are propagated in both time directions: one simulation runs forward and one simulation runs backward in time. During the simulations, the nucleus size $n$ is calculated every $250$ timesteps according to a criterion based on `solid-like bonds' as explained in \Cref{sec:solid-like}.
The simulation is terminated when the nucleus size $n<30$ or when $n>2000$.
The new trajectory is accepted if it  shows successful nucleation, i.e. if one of the simulations ends with $n<30$ and the other with $n>2000$. 
To be clear, there are four possible outcomes: (1) the forward path ends with $n<30$ and the backward path with $n>2000$; (2) the backward path ends with $n<30$ and the forward path with $n>2000$; (3) both paths end with $n<30$; (4) both paths end with $n>2000$.
The new trajectory is accepted in the first two cases and rejected in the last two cases.


To select the shooting points, we use the following procedure that is slightly adapted from Ref.\ \citenum{peters_obtaining_2006}. From the initial path, we randomly pick  shooting points until a trajectory is accepted. Subsequently, we choose a new shooting point $t_n$  by slightly perturbing a previously accepted shooting point $t_o$: The new shooting point $t_n$ is randomly selected  from one of the following candidates:
$t_o, t_o\pm 1 \Delta T, t_o\pm 2 \Delta T, \dots, t_o\pm 20 \Delta T$.
where $\Delta T = \sqrt{m \sigma^2 / \epsilon}$ should not be confused with the simulation timestep $\Delta t = 0.004 \sqrt{m \sigma^2 / \epsilon}$.
The typical length of accepted trajectories is around $650 \Delta T$ with a standard deviation of $80 \Delta T$, so the time interval $|t_n - t_o|$ is much smaller than the total length of the trajectory.

\begin{figure}
    \centering
    \includegraphics[width=\linewidth]{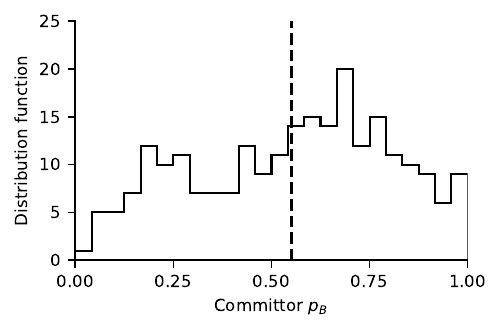}
    \caption{Distribution of committor values $\hat{p}_B$ computed for $240$ different shooting points. The distribution has a mean value of $0.55$ visualized by the dashed black line and a standard deviation of $0.25$.}
    \label{fig:committor-distribution}
\end{figure}

Since shooting points are only accepted if they lead to a transition path, the nuclei at these shooting points must have a reasonable probability of both growing and melting. By choosing  new shooting points close to  previously accepted ones, aimless shooting leads to shooting points that are close to the transition states.\cite{peters_obtaining_2006} 
To identify critical nuclei, we  selected $240$ different shooting points and  performed a committor analysis on them. For each shooting point, we performed $M=100$ different simulations to approximate the committor $\hat{p}_B$, i.e.\ the probability that the nucleus in that  shooting point will grow rather than melt.
In \Cref{fig:committor-distribution}, we present the distribution of committor values $\hat{p}_B$ for the $240$ shooting points.
We see that the committor distribution is broad, with a standard deviation of $0.25$ and a mean value of $0.55$, indicating that the shooting points are not guaranteed to be transition states.
We deem a nucleus critical if the approximated committor $\hat{p}_B$ is statistically indistinguishable from $0.5$, i.e.\ if $0.5$ lies within the interval $[\hat{p}_B-2\sigma, \hat{p}_B+2\sigma]$, where $\sigma = \sqrt{\hat{p}_B(1-\hat{p}_B) / M}$.\cite{scholl-paschinger_demixing_2010}
Using this criterion, we identified $53$ critical nuclei from the $240$ analyzed shooting points. 
These $53$ critical nuclei are the ones we use for further analysis in the main text.

We use a total of $10,000$ shooting moves with an average acceptance rate of $24\%$. We use the first $5,000$ shooting moves for equilibration and the last $5,000$ for production. At each shooting point, we use  polyhedral template matching,\cite{larsen_robust_2016} to determine the number of fcc-like  $n_{\mathrm{fcc}}$, hcp-like $n_{\mathrm{hcp}}$ and bcc-like $n_{\mathrm{bcc}}$ particles in the nucleus. We use this data to calculate the autocorrelation function of the fraction $n_{\mathrm{fcc}} / (n_{\mathrm{fcc}}+n_{\mathrm{hcp}}+n_{\mathrm{bcc}})$ of fcc-like particles. Using this autocorrelation function, we estimate the decorrelation time to be around $100$ shooting moves. 
Therefore, the pathways we used for further structural analysis were obtained after every $100$ shooting moves. 
This results in a total of $50$ decorrelated nucleation pathways.

Recently, \citet{falkner_revisiting_2024} demonstrated that the aimless shooting procedure with flexible path lengths, as  described above, introduces a bias in the transition path ensemble.
To correct this bias, they proposed  a reweighting procedure where  each transition path is assigned a relative weight of $1/L$ where $L$ is the length of the path. 
We have incorporated this reweighting procedure into our analysis: each critical nucleus or nucleation trajectory used in \Cref{fig:compositions} has been reweighted to calculate the average compositions.

To assess the impact of reweighting, we  compared the results  in \Cref{fig:compositions} both with and without reweighting. 
We found that the effect of reweighting is minimal in our case and is not visible in \Cref{fig:compositions}. 
Ref.\ \citenum{falkner_revisiting_2024} notes that the bias in the transition path ensemble is less significant for higher free-energy barriers, indicating that the minor impact of reweighting in our case may  result from the fact that the nucleation barrier is around $25~k_B T$,\cite{ten_wolde_numerical_1995} significantly higher than the $5~k_B T$ barrier investigated in Ref.\ \citenum{falkner_revisiting_2024}.




\section{Bond orientational order parameters}
\label{sec:bops}
We used the freud library \cite{ramasubramani_freud_2020} to calculate the local bond order parameters.
To calculate the local bond orientational order parameters for a particle $i$, we first identify its nearest neighbors. In most cases, this was done using the solid-angle-based nearest neighbor algorithm (SANN).\cite{van_meel_parameter-free_2012} For the machine-learning schemes, we also determined nearest neighbors by selecting the $k$ particles which are closest to particle $i$.

In addition to this definition of nearest neighbors, we follow the definition of $d_6 (i,j)$ given in Ref.\ \citenum{filion_crystal_2010}, and the definitions of $q_l$, $w_l$, $\overline{q}_l$ and $\overline{w}_l$ given in Ref.\ \citenum{lechner_accurate_2008}.

\section{Neural network feature importance}
\label{sec:sensitivity}

In this Appendix, we perform a sensitivity analysis to identify the most important features out of  the $195$ input features for the ML1 and ML2 schemes in polymorph classification. To do this, we apply the permutation sensitivity analysis.\cite{scikit-learn, breiman_random_2001}
For each feature, we measure the decrease in classification accuracy on the test set, when the values of that feature are randomly permuted. The reduction in classification accuracy serves as  a measure of the feature's importance to the classifier.

\begin{figure}
    \centering
    \includegraphics[width=0.95\linewidth]{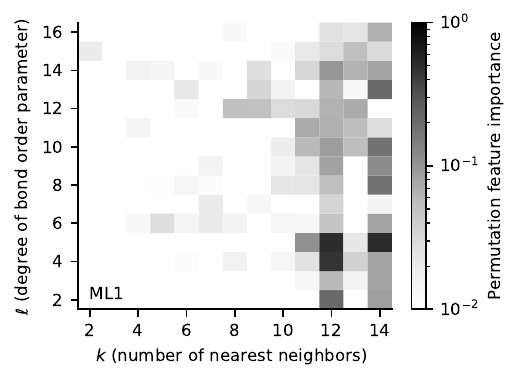} \\
    \includegraphics[width=0.95\linewidth]{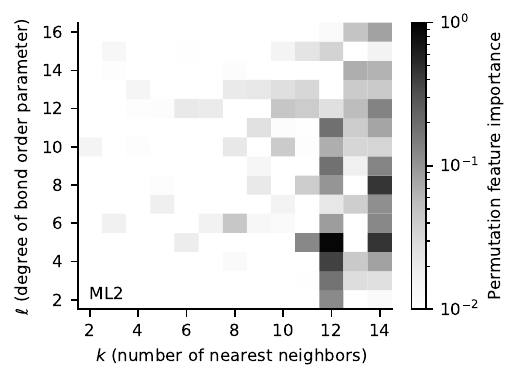}
    \caption{Feature importance of the $195$ input features of the neural networks used in the ML1 and ML2 schemes for polymorph classification.
    The $195$ features include $15$ non-averaged bond order parameters $q_l$ for each number of nearest neighbors $k$ ranging from $2$ to $14$.
    The feature importance is determined using a permutation sensitivity analysis.\cite{scikit-learn, breiman_random_2001}}
    \label{fig:feature-importance}
\end{figure}

In \Cref{fig:feature-importance}, we present the permutation feature importance for all $195$ features used in the ML1 and ML2 schemes. 
These $195$ features include $15$ non-averaged bond order parameters $q_l$ for each number of nearest neighbors $k$ ranging from $2$ to $14$, as described in the main text.
The feature importance is shown in  grayscale according to the colorbar on the right of the plots. More important features are shown in darker shades, reflecting a significant decrease in classification accuracy.

For both ML1 and ML2, we observe that the most important features correspond to $k=12$ and $k=14$ nearest neighbors, which  makes sense given that  fcc and hcp have 12 nearest neighbours, and bcc has $14$ nearest and next-nearest neighbors.
Features with $k=8,9,10,11,13$ are also relatively important. Bond order parameters of almost all degrees $l$ contribute to the classification, but $l=5$ appears particularly important for the classification problem.
Overall, we see that ML1 and ML2 rely on a similar set of features, with a strong dependence on $k\geq 8$. 
If we define a feature as `relevant' when the decrease in classification accuracy exceeds $1\%$, we find that  slightly less than half of the features are relevant to the classification problem.





\section{Fluctuations and stability of the bcc phase}
\label{sec:bcc-stability}
The bcc crystal phase is known to be mechanically unstable for  Lennard-Jones particles.\cite{schwerdtfeger_cuboidal_2022} The mechanical instability is easily observed when using an anisotropic barostat, as in this case, the bcc phase  quickly transitions to a mixture of fcc and hcp structures. In contrast, with an isotropic barostat or in the canonical ($NVT$) ensemble, the bcc phase seems metastable. \citet{eshet_new_2008} demonstrated that  simulations with an isotropic barostat can  result in a new metastable crystal phase. They refer to this new phase as the $I\overline{4}3d$ or `distorted bcc' phase. However, we have not observed significant differences in the stability between bcc or $I\overline{4}3d$ at our simulation conditions: both phases rapidly  convert, almost instantaneously, to a mixture of fcc and hcp when using an anisotropic barostat.

\begin{figure}
    \centering
    \includegraphics[width=0.95\linewidth]{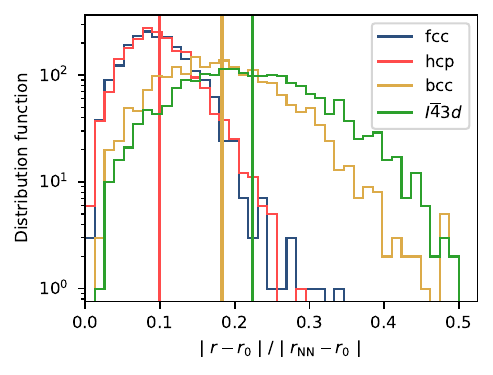}
    \caption{Distribution of distances from ideal lattice positions for thermally equilibrated fcc, hcp, bcc and $I\overline{4}3d$ phases at our simulation conditions. Distances are normalized by the nearest neighbor distance in the ideal lattices. The vertical lines correspond to the average distance for each phase.}
    \label{fig:fluctuations}
\end{figure}

When using an isotropic barostat, we  observe that both bcc and $I\overline{4}3d$ display significantly larger fluctuations around their ideal lattice positions compared to fcc and hcp phases. In \Cref{fig:fluctuations}, we show the distribution of distances from the ideal lattice positions for bulk crystal phases  under our simulation conditions when using an isotropic barostat. The mean distance from the ideal positions is shown as a vertical line for each phase. For the fcc and hcp phases, the mean distance is around $10\%$ of the nearest neighbor distance, around $20\%$ for bcc, and as large as $24\%$ for $I\overline{4}3d$. These large fluctuations naturally lead to large fluctuations in the bond orientational order parameters.

Ref.\ \citenum{jungblut_crystallization_2011} suggested that the wide $\w_4$-distribution of a thermally equilibrated bcc phase points to fluctuations between bcc and $I\overline{4}3d$. When  applying polyhedral template matching to the bcc phase equilibrated with an isotropic barostat (with a RMSD cutoff of $0.12$), we find that around $50\%$ of the particles are classified as locally bcc-like. Smaller fractions of fcc and hcp particles,  around $5-10\%$ each, are also identified, while  around $40\%$ are not classified as fcc, hcp or bcc. We obtain similar results with our ML1 scheme. Thus, the bcc phase equilibrated with an isotropic barostat appears to correspond either to the $I\overline{4}3d$ phase or a heterogeneous mixture of fcc, hcp, bcc and $I\overline{4}3d$. Further structural characterization of this state would be an interesting topic for future research.
Unfortunately, none of the local structure detection schemes we employed in this paper were specifically designed to identify the  $I\overline{4}3d$ phase.
Although our ML1 scheme could be extended to include the $I\overline{4}3d$ phase, it would be impossible to compare the resulting classification with any of the other methods. 

\bibliography{main}%

\begin{thebibliography}{71}%
\makeatletter
\providecommand \@ifxundefined [1]{%
 \@ifx{#1\undefined}
}%
\providecommand \@ifnum [1]{%
 \ifnum #1\expandafter \@firstoftwo
 \else \expandafter \@secondoftwo
 \fi
}%
\providecommand \@ifx [1]{%
 \ifx #1\expandafter \@firstoftwo
 \else \expandafter \@secondoftwo
 \fi
}%
\providecommand \natexlab [1]{#1}%
\providecommand \enquote  [1]{``#1''}%
\providecommand \bibnamefont  [1]{#1}%
\providecommand \bibfnamefont [1]{#1}%
\providecommand \citenamefont [1]{#1}%
\providecommand \href@noop [0]{\@secondoftwo}%
\providecommand \href [0]{\begingroup \@sanitize@url \@href}%
\providecommand \@href[1]{\@@startlink{#1}\@@href}%
\providecommand \@@href[1]{\endgroup#1\@@endlink}%
\providecommand \@sanitize@url [0]{\catcode `\\12\catcode `\$12\catcode `\&12\catcode `\#12\catcode `\^12\catcode `\_12\catcode `\%12\relax}%
\providecommand \@@startlink[1]{}%
\providecommand \@@endlink[0]{}%
\providecommand \url  [0]{\begingroup\@sanitize@url \@url }%
\providecommand \@url [1]{\endgroup\@href {#1}{\urlprefix }}%
\providecommand \urlprefix  [0]{URL }%
\providecommand \Eprint [0]{\href }%
\providecommand \doibase [0]{http://dx.doi.org/}%
\providecommand \selectlanguage [0]{\@gobble}%
\providecommand \bibinfo  [0]{\@secondoftwo}%
\providecommand \bibfield  [0]{\@secondoftwo}%
\providecommand \translation [1]{[#1]}%
\providecommand \BibitemOpen [0]{}%
\providecommand \bibitemStop [0]{}%
\providecommand \bibitemNoStop [0]{.\EOS\space}%
\providecommand \EOS [0]{\spacefactor3000\relax}%
\providecommand \BibitemShut  [1]{\csname bibitem#1\endcsname}%
\let\auto@bib@innerbib\@empty
\bibitem [{\citenamefont {ten Wolde}, \citenamefont {Ruiz-Montero},\ and\ \citenamefont {Frenkel}(1995)}]{ten_wolde_numerical_1995}%
  \BibitemOpen
  \bibfield  {author} {\bibinfo {author} {\bibfnamefont {P.~R.}\ \bibnamefont {ten Wolde}}, \bibinfo {author} {\bibfnamefont {M.~J.}\ \bibnamefont {Ruiz-Montero}}, \ and\ \bibinfo {author} {\bibfnamefont {D.}~\bibnamefont {Frenkel}},\ }\bibfield  {title} {\enquote {\bibinfo {title} {Numerical {Evidence} for bcc {Ordering} at the {Surface} of a {Critical} fcc {Nucleus}},}\ }\href {\doibase 10.1103/PhysRevLett.75.2714} {\bibfield  {journal} {\bibinfo  {journal} {Phys. Rev. Lett.}\ }\textbf {\bibinfo {volume} {75}},\ \bibinfo {pages} {2714--2717} (\bibinfo {year} {1995})}\BibitemShut {NoStop}%
\bibitem [{\citenamefont {Stranski}\ and\ \citenamefont {Totomanow}(1933)}]{stranski_rate_1933}%
  \BibitemOpen
  \bibfield  {author} {\bibinfo {author} {\bibfnamefont {I.}~\bibnamefont {Stranski}}\ and\ \bibinfo {author} {\bibfnamefont {D.}~\bibnamefont {Totomanow}},\ }\bibfield  {title} {\enquote {\bibinfo {title} {Rate of formation of (crystal) nuclei and the ostwald step rule},}\ }\href@noop {} {\bibfield  {journal} {\bibinfo  {journal} {Z. Phys. Chem}\ }\textbf {\bibinfo {volume} {163}},\ \bibinfo {pages} {399--408} (\bibinfo {year} {1933})}\BibitemShut {NoStop}%
\bibitem [{\citenamefont {Alexander}\ and\ \citenamefont {McTague}(1978)}]{alexander_should_1978}%
  \BibitemOpen
  \bibfield  {author} {\bibinfo {author} {\bibfnamefont {S.}~\bibnamefont {Alexander}}\ and\ \bibinfo {author} {\bibfnamefont {J.}~\bibnamefont {McTague}},\ }\bibfield  {title} {\enquote {\bibinfo {title} {Should {All} {Crystals} {Be} bcc? {Landau} {Theory} of {Solidification} and {Crystal} {Nucleation}},}\ }\href {\doibase 10.1103/PhysRevLett.41.702} {\bibfield  {journal} {\bibinfo  {journal} {Phys. Rev. Lett.}\ }\textbf {\bibinfo {volume} {41}},\ \bibinfo {pages} {702--705} (\bibinfo {year} {1978})}\BibitemShut {NoStop}%
\bibitem [{\citenamefont {Fortini}, \citenamefont {Sanz},\ and\ \citenamefont {Dijkstra}(2008)}]{fortini_crystallization_2008}%
  \BibitemOpen
  \bibfield  {author} {\bibinfo {author} {\bibfnamefont {A.}~\bibnamefont {Fortini}}, \bibinfo {author} {\bibfnamefont {E.}~\bibnamefont {Sanz}}, \ and\ \bibinfo {author} {\bibfnamefont {M.}~\bibnamefont {Dijkstra}},\ }\bibfield  {title} {\enquote {\bibinfo {title} {Crystallization and gelation in colloidal systems with short-ranged attractive interactions},}\ }\href {\doibase 10.1103/PhysRevE.78.041402} {\bibfield  {journal} {\bibinfo  {journal} {Phys. Rev. E}\ }\textbf {\bibinfo {volume} {78}},\ \bibinfo {pages} {041402} (\bibinfo {year} {2008})}\BibitemShut {NoStop}%
\bibitem [{\citenamefont {Ji}\ \emph {et~al.}(2018)\citenamefont {Ji}, \citenamefont {Sun}, \citenamefont {Ouyang},\ and\ \citenamefont {Xu}}]{ji_crystal_2018}%
  \BibitemOpen
  \bibfield  {author} {\bibinfo {author} {\bibfnamefont {X.}~\bibnamefont {Ji}}, \bibinfo {author} {\bibfnamefont {Z.}~\bibnamefont {Sun}}, \bibinfo {author} {\bibfnamefont {W.}~\bibnamefont {Ouyang}}, \ and\ \bibinfo {author} {\bibfnamefont {S.}~\bibnamefont {Xu}},\ }\bibfield  {title} {\enquote {\bibinfo {title} {Crystal nucleation and metastable bcc phase in charged colloids: {A} molecular dynamics study},}\ }\href {\doibase 10.1063/1.5016235} {\bibfield  {journal} {\bibinfo  {journal} {J. Chem. Phys.}\ }\textbf {\bibinfo {volume} {148}},\ \bibinfo {pages} {174904} (\bibinfo {year} {2018})}\BibitemShut {NoStop}%
\bibitem [{\citenamefont {Kratzer}\ and\ \citenamefont {Arnold}(2015)}]{kratzer_two-stage_2015}%
  \BibitemOpen
  \bibfield  {author} {\bibinfo {author} {\bibfnamefont {K.}~\bibnamefont {Kratzer}}\ and\ \bibinfo {author} {\bibfnamefont {A.}~\bibnamefont {Arnold}},\ }\bibfield  {title} {\enquote {\bibinfo {title} {Two-stage crystallization of charged colloids under low supersaturation conditions},}\ }\href {\doibase 10.1039/C4SM02365J} {\bibfield  {journal} {\bibinfo  {journal} {Soft Matter}\ }\textbf {\bibinfo {volume} {11}},\ \bibinfo {pages} {2174--2182} (\bibinfo {year} {2015})}\BibitemShut {NoStop}%
\bibitem [{\citenamefont {Gispen}\ and\ \citenamefont {Dijkstra}(2022)}]{gispen_kinetic_2022}%
  \BibitemOpen
  \bibfield  {author} {\bibinfo {author} {\bibfnamefont {W.}~\bibnamefont {Gispen}}\ and\ \bibinfo {author} {\bibfnamefont {M.}~\bibnamefont {Dijkstra}},\ }\bibfield  {title} {\enquote {\bibinfo {title} {Kinetic {Phase} {Diagram} for {Nucleation} and {Growth} of {Competing} {Crystal} {Polymorphs} in {Charged} {Colloids}},}\ }\href {\doibase https://doi.org/10.1103/PhysRevLett.129.098002} {\bibfield  {journal} {\bibinfo  {journal} {Phys. Rev. Lett.}\ }\textbf {\bibinfo {volume} {129}},\ \bibinfo {pages} {098002} (\bibinfo {year} {2022})}\BibitemShut {NoStop}%
\bibitem [{\citenamefont {Sadigh}, \citenamefont {Zepeda-Ruiz},\ and\ \citenamefont {Belof}(2021)}]{sadigh_metastablesolid_2021}%
  \BibitemOpen
  \bibfield  {author} {\bibinfo {author} {\bibfnamefont {B.}~\bibnamefont {Sadigh}}, \bibinfo {author} {\bibfnamefont {L.}~\bibnamefont {Zepeda-Ruiz}}, \ and\ \bibinfo {author} {\bibfnamefont {J.~L.}\ \bibnamefont {Belof}},\ }\bibfield  {title} {\enquote {\bibinfo {title} {Metastable–solid phase diagrams derived from polymorphic solidification kinetics},}\ }\href {\doibase 10.1073/pnas.2017809118} {\bibfield  {journal} {\bibinfo  {journal} {Proc. Natl. Acad. Sci. U.S.A.}\ }\textbf {\bibinfo {volume} {118}},\ \bibinfo {pages} {e2017809118} (\bibinfo {year} {2021})}\BibitemShut {NoStop}%
\bibitem [{\citenamefont {Sun}\ \emph {et~al.}(2022)\citenamefont {Sun}, \citenamefont {Zhang}, \citenamefont {Mendelev}, \citenamefont {Wentzcovitch},\ and\ \citenamefont {Ho}}]{sun_two-step_2022}%
  \BibitemOpen
  \bibfield  {author} {\bibinfo {author} {\bibfnamefont {Y.}~\bibnamefont {Sun}}, \bibinfo {author} {\bibfnamefont {F.}~\bibnamefont {Zhang}}, \bibinfo {author} {\bibfnamefont {M.~I.}\ \bibnamefont {Mendelev}}, \bibinfo {author} {\bibfnamefont {R.~M.}\ \bibnamefont {Wentzcovitch}}, \ and\ \bibinfo {author} {\bibfnamefont {K.-M.}\ \bibnamefont {Ho}},\ }\bibfield  {title} {\enquote {\bibinfo {title} {Two-step nucleation of the {Earth}’s inner core},}\ }\href {\doibase 10.1073/pnas.2113059119} {\bibfield  {journal} {\bibinfo  {journal} {Proc. Natl. Acad. Sci. U.S.A.}\ }\textbf {\bibinfo {volume} {119}},\ \bibinfo {pages} {e2113059119} (\bibinfo {year} {2022})}\BibitemShut {NoStop}%
\bibitem [{\citenamefont {{Arjun}}, \citenamefont {Berendsen},\ and\ \citenamefont {Bolhuis}(2019)}]{arjun_unbiased_2019}%
  \BibitemOpen
  \bibfield  {author} {\bibinfo {author} {\bibnamefont {{Arjun}}}, \bibinfo {author} {\bibfnamefont {T.~A.}\ \bibnamefont {Berendsen}}, \ and\ \bibinfo {author} {\bibfnamefont {P.~G.}\ \bibnamefont {Bolhuis}},\ }\bibfield  {title} {\enquote {\bibinfo {title} {Unbiased atomistic insight in the competing nucleation mechanisms of methane hydrates},}\ }\href {\doibase 10.1073/pnas.1906502116} {\bibfield  {journal} {\bibinfo  {journal} {Proc. Natl. Acad. Sci. U.S.A.}\ }\textbf {\bibinfo {volume} {116}},\ \bibinfo {pages} {19305--19310} (\bibinfo {year} {2019})}\BibitemShut {NoStop}%
\bibitem [{\citenamefont {Bulutoglu}\ \emph {et~al.}(2022)\citenamefont {Bulutoglu}, \citenamefont {Wang}, \citenamefont {Boukerche}, \citenamefont {Nere}, \citenamefont {Corti},\ and\ \citenamefont {Ramkrishna}}]{bulutoglu_investigation_2022}%
  \BibitemOpen
  \bibfield  {author} {\bibinfo {author} {\bibfnamefont {P.~S.}\ \bibnamefont {Bulutoglu}}, \bibinfo {author} {\bibfnamefont {S.}~\bibnamefont {Wang}}, \bibinfo {author} {\bibfnamefont {M.}~\bibnamefont {Boukerche}}, \bibinfo {author} {\bibfnamefont {N.~K.}\ \bibnamefont {Nere}}, \bibinfo {author} {\bibfnamefont {D.~S.}\ \bibnamefont {Corti}}, \ and\ \bibinfo {author} {\bibfnamefont {D.}~\bibnamefont {Ramkrishna}},\ }\bibfield  {title} {\enquote {\bibinfo {title} {An investigation of the kinetics and thermodynamics of {NaCl} nucleation through composite clusters},}\ }\href {\doibase 10.1093/pnasnexus/pgac033} {\bibfield  {journal} {\bibinfo  {journal} {PNAS Nexus}\ }\textbf {\bibinfo {volume} {1}},\ \bibinfo {pages} {pgac033} (\bibinfo {year} {2022})}\BibitemShut {NoStop}%
\bibitem [{\citenamefont {Desgranges}\ and\ \citenamefont {Delhommelle}(2007)}]{desgranges_controlling_2007}%
  \BibitemOpen
  \bibfield  {author} {\bibinfo {author} {\bibfnamefont {C.}~\bibnamefont {Desgranges}}\ and\ \bibinfo {author} {\bibfnamefont {J.}~\bibnamefont {Delhommelle}},\ }\bibfield  {title} {\enquote {\bibinfo {title} {Controlling {Polymorphism} during the {Crystallization} of an {Atomic} {Fluid}},}\ }\href {\doibase 10.1103/PhysRevLett.98.235502} {\bibfield  {journal} {\bibinfo  {journal} {Phys. Rev. Lett.}\ }\textbf {\bibinfo {volume} {98}},\ \bibinfo {pages} {235502} (\bibinfo {year} {2007})}\BibitemShut {NoStop}%
\bibitem [{\citenamefont {van Meel}\ \emph {et~al.}(2008)\citenamefont {van Meel}, \citenamefont {Page}, \citenamefont {Sear},\ and\ \citenamefont {Frenkel}}]{van_meel_two-step_2008}%
  \BibitemOpen
  \bibfield  {author} {\bibinfo {author} {\bibfnamefont {J.~A.}\ \bibnamefont {van Meel}}, \bibinfo {author} {\bibfnamefont {A.~J.}\ \bibnamefont {Page}}, \bibinfo {author} {\bibfnamefont {R.~P.}\ \bibnamefont {Sear}}, \ and\ \bibinfo {author} {\bibfnamefont {D.}~\bibnamefont {Frenkel}},\ }\bibfield  {title} {\enquote {\bibinfo {title} {Two-step vapor-crystal nucleation close below triple point},}\ }\href {\doibase 10.1063/1.3026364} {\bibfield  {journal} {\bibinfo  {journal} {J. Chem. Phys.}\ }\textbf {\bibinfo {volume} {129}},\ \bibinfo {pages} {204505} (\bibinfo {year} {2008})}\BibitemShut {NoStop}%
\bibitem [{\citenamefont {Shen}\ and\ \citenamefont {Oxtoby}(1996)}]{shen_bcc_1996}%
  \BibitemOpen
  \bibfield  {author} {\bibinfo {author} {\bibfnamefont {Y.~C.}\ \bibnamefont {Shen}}\ and\ \bibinfo {author} {\bibfnamefont {D.~W.}\ \bibnamefont {Oxtoby}},\ }\bibfield  {title} {\enquote {\bibinfo {title} {bcc {Symmetry} in the {Crystal}-{Melt} {Interface} of {Lennard}-{Jones} {Fluids} {Examined} through {Density} {Functional} {Theory}},}\ }\href {\doibase 10.1103/PhysRevLett.77.3585} {\bibfield  {journal} {\bibinfo  {journal} {Phys. Rev. Lett.}\ }\textbf {\bibinfo {volume} {77}},\ \bibinfo {pages} {3585--3588} (\bibinfo {year} {1996})}\BibitemShut {NoStop}%
\bibitem [{\citenamefont {Wang}, \citenamefont {Mi},\ and\ \citenamefont {Zhong}(2013)}]{wang_density_2013}%
  \BibitemOpen
  \bibfield  {author} {\bibinfo {author} {\bibfnamefont {X.}~\bibnamefont {Wang}}, \bibinfo {author} {\bibfnamefont {J.}~\bibnamefont {Mi}}, \ and\ \bibinfo {author} {\bibfnamefont {C.}~\bibnamefont {Zhong}},\ }\bibfield  {title} {\enquote {\bibinfo {title} {Density functional theory for crystal-liquid interfaces of {Lennard}-{Jones} fluid},}\ }\href {\doibase 10.1063/1.4802633} {\bibfield  {journal} {\bibinfo  {journal} {J. Chem. Phys.}\ }\textbf {\bibinfo {volume} {138}},\ \bibinfo {pages} {164704} (\bibinfo {year} {2013})}\BibitemShut {NoStop}%
\bibitem [{\citenamefont {Schoonen}\ and\ \citenamefont {Lutsko}(2022)}]{schoonen_crystal_2022}%
  \BibitemOpen
  \bibfield  {author} {\bibinfo {author} {\bibfnamefont {C.}~\bibnamefont {Schoonen}}\ and\ \bibinfo {author} {\bibfnamefont {J.~F.}\ \bibnamefont {Lutsko}},\ }\bibfield  {title} {\enquote {\bibinfo {title} {Crystal {Polymorphism} {Induced} by {Surface} {Tension}},}\ }\href {\doibase 10.1103/PhysRevLett.129.246101} {\bibfield  {journal} {\bibinfo  {journal} {Phys. Rev. Lett.}\ }\textbf {\bibinfo {volume} {129}},\ \bibinfo {pages} {246101} (\bibinfo {year} {2022})}\BibitemShut {NoStop}%
\bibitem [{\citenamefont {Moroni}, \citenamefont {ten Wolde},\ and\ \citenamefont {Bolhuis}(2005)}]{moroni_interplay_2005}%
  \BibitemOpen
  \bibfield  {author} {\bibinfo {author} {\bibfnamefont {D.}~\bibnamefont {Moroni}}, \bibinfo {author} {\bibfnamefont {P.~R.}\ \bibnamefont {ten Wolde}}, \ and\ \bibinfo {author} {\bibfnamefont {P.~G.}\ \bibnamefont {Bolhuis}},\ }\bibfield  {title} {\enquote {\bibinfo {title} {Interplay between {Structure} and {Size} in a {Critical} {Crystal} {Nucleus}},}\ }\href {\doibase 10.1103/PhysRevLett.94.235703} {\bibfield  {journal} {\bibinfo  {journal} {Phys. Rev. Lett.}\ }\textbf {\bibinfo {volume} {94}},\ \bibinfo {pages} {235703} (\bibinfo {year} {2005})}\BibitemShut {NoStop}%
\bibitem [{\citenamefont {Eslami}, \citenamefont {Khanjari},\ and\ \citenamefont {Müller-Plathe}(2017)}]{eslami_local_2017}%
  \BibitemOpen
  \bibfield  {author} {\bibinfo {author} {\bibfnamefont {H.}~\bibnamefont {Eslami}}, \bibinfo {author} {\bibfnamefont {N.}~\bibnamefont {Khanjari}}, \ and\ \bibinfo {author} {\bibfnamefont {F.}~\bibnamefont {Müller-Plathe}},\ }\bibfield  {title} {\enquote {\bibinfo {title} {A {Local} {Order} {Parameter}-{Based} {Method} for {Simulation} of {Free} {Energy} {Barriers} in {Crystal} {Nucleation}},}\ }\href {\doibase 10.1021/acs.jctc.6b01034} {\bibfield  {journal} {\bibinfo  {journal} {J. Chem. Theory Comput.}\ }\textbf {\bibinfo {volume} {13}},\ \bibinfo {pages} {1307--1316} (\bibinfo {year} {2017})}\BibitemShut {NoStop}%
\bibitem [{\citenamefont {Prestipino}(2018)}]{prestipino_barrier_2018}%
  \BibitemOpen
  \bibfield  {author} {\bibinfo {author} {\bibfnamefont {S.}~\bibnamefont {Prestipino}},\ }\bibfield  {title} {\enquote {\bibinfo {title} {The barrier to ice nucleation in monatomic water},}\ }\href {\doibase 10.1063/1.5016518} {\bibfield  {journal} {\bibinfo  {journal} {J. Chem. Phys.}\ }\textbf {\bibinfo {volume} {148}},\ \bibinfo {pages} {124505} (\bibinfo {year} {2018})}\BibitemShut {NoStop}%
\bibitem [{\citenamefont {Jungblut}\ and\ \citenamefont {Dellago}(2011)}]{jungblut_crystallization_2011}%
  \BibitemOpen
  \bibfield  {author} {\bibinfo {author} {\bibfnamefont {S.}~\bibnamefont {Jungblut}}\ and\ \bibinfo {author} {\bibfnamefont {C.}~\bibnamefont {Dellago}},\ }\bibfield  {title} {\enquote {\bibinfo {title} {Crystallization of a binary {Lennard}-{Jones} mixture},}\ }\href {\doibase 10.1063/1.3556664} {\bibfield  {journal} {\bibinfo  {journal} {J. Chem. Phys.}\ }\textbf {\bibinfo {volume} {134}},\ \bibinfo {pages} {104501} (\bibinfo {year} {2011})}\BibitemShut {NoStop}%
\bibitem [{\citenamefont {Ouyang}\ \emph {et~al.}(2020)\citenamefont {Ouyang}, \citenamefont {Sun}, \citenamefont {Sun},\ and\ \citenamefont {Xu}}]{ouyang_entire_2020}%
  \BibitemOpen
  \bibfield  {author} {\bibinfo {author} {\bibfnamefont {W.}~\bibnamefont {Ouyang}}, \bibinfo {author} {\bibfnamefont {B.}~\bibnamefont {Sun}}, \bibinfo {author} {\bibfnamefont {Z.}~\bibnamefont {Sun}}, \ and\ \bibinfo {author} {\bibfnamefont {S.}~\bibnamefont {Xu}},\ }\bibfield  {title} {\enquote {\bibinfo {title} {Entire crystallization process of {Lennard}-{Jones} liquids: {A} large-scale molecular dynamics study},}\ }\href {\doibase 10.1063/1.5139574} {\bibfield  {journal} {\bibinfo  {journal} {J. Chem. Phys.}\ }\textbf {\bibinfo {volume} {152}},\ \bibinfo {pages} {054903} (\bibinfo {year} {2020})}\BibitemShut {NoStop}%
\bibitem [{\citenamefont {Russo}, \citenamefont {Romano},\ and\ \citenamefont {Tanaka}(2014)}]{russo_new_2014}%
  \BibitemOpen
  \bibfield  {author} {\bibinfo {author} {\bibfnamefont {J.}~\bibnamefont {Russo}}, \bibinfo {author} {\bibfnamefont {F.}~\bibnamefont {Romano}}, \ and\ \bibinfo {author} {\bibfnamefont {H.}~\bibnamefont {Tanaka}},\ }\bibfield  {title} {\enquote {\bibinfo {title} {New metastable form of ice and its role in the homogeneous crystallization of water},}\ }\href {\doibase 10.1038/nmat3977} {\bibfield  {journal} {\bibinfo  {journal} {Nat. Mater.}\ }\textbf {\bibinfo {volume} {13}},\ \bibinfo {pages} {733--739} (\bibinfo {year} {2014})}\BibitemShut {NoStop}%
\bibitem [{\citenamefont {Espinosa}\ \emph {et~al.}(2016{\natexlab{a}})\citenamefont {Espinosa}, \citenamefont {Zaragoza}, \citenamefont {Rosales-Pelaez}, \citenamefont {Navarro}, \citenamefont {Valeriani}, \citenamefont {Vega},\ and\ \citenamefont {Sanz}}]{espinosa_interfacial_2016}%
  \BibitemOpen
  \bibfield  {author} {\bibinfo {author} {\bibfnamefont {J.~R.}\ \bibnamefont {Espinosa}}, \bibinfo {author} {\bibfnamefont {A.}~\bibnamefont {Zaragoza}}, \bibinfo {author} {\bibfnamefont {P.}~\bibnamefont {Rosales-Pelaez}}, \bibinfo {author} {\bibfnamefont {C.}~\bibnamefont {Navarro}}, \bibinfo {author} {\bibfnamefont {C.}~\bibnamefont {Valeriani}}, \bibinfo {author} {\bibfnamefont {C.}~\bibnamefont {Vega}}, \ and\ \bibinfo {author} {\bibfnamefont {E.}~\bibnamefont {Sanz}},\ }\bibfield  {title} {\enquote {\bibinfo {title} {Interfacial {Free} {Energy} as the {Key} to the {Pressure}-{Induced} {Deceleration} of {Ice} {Nucleation}},}\ }\href {\doibase 10.1103/PhysRevLett.117.135702} {\bibfield  {journal} {\bibinfo  {journal} {Phys. Rev. Lett.}\ }\textbf {\bibinfo {volume} {117}},\ \bibinfo {pages} {135702} (\bibinfo {year} {2016}{\natexlab{a}})}\BibitemShut {NoStop}%
\bibitem [{\citenamefont {Leoni}\ and\ \citenamefont {Russo}(2021)}]{leoni_nonclassical_2021}%
  \BibitemOpen
  \bibfield  {author} {\bibinfo {author} {\bibfnamefont {F.}~\bibnamefont {Leoni}}\ and\ \bibinfo {author} {\bibfnamefont {J.}~\bibnamefont {Russo}},\ }\bibfield  {title} {\enquote {\bibinfo {title} {Nonclassical {Nucleation} {Pathways} in {Stacking}-{Disordered} {Crystals}},}\ }\href {\doibase 10.1103/PhysRevX.11.031006} {\bibfield  {journal} {\bibinfo  {journal} {Phys. Rev. X}\ }\textbf {\bibinfo {volume} {11}},\ \bibinfo {pages} {031006} (\bibinfo {year} {2021})}\BibitemShut {NoStop}%
\bibitem [{\citenamefont {Kawasaki}\ and\ \citenamefont {Tanaka}(2010)}]{kawasaki_formation_2010}%
  \BibitemOpen
  \bibfield  {author} {\bibinfo {author} {\bibfnamefont {T.}~\bibnamefont {Kawasaki}}\ and\ \bibinfo {author} {\bibfnamefont {H.}~\bibnamefont {Tanaka}},\ }\bibfield  {title} {\enquote {\bibinfo {title} {Formation of a crystal nucleus from liquid},}\ }\href {\doibase 10.1073/pnas.1001040107} {\bibfield  {journal} {\bibinfo  {journal} {Proc. Natl. Acad. Sci. U.S.A.}\ }\textbf {\bibinfo {volume} {107}} (\bibinfo {year} {2010}),\ 10.1073/pnas.1001040107}\BibitemShut {NoStop}%
\bibitem [{\citenamefont {Lechner}, \citenamefont {Dellago},\ and\ \citenamefont {Bolhuis}(2011)}]{lechner_role_2011}%
  \BibitemOpen
  \bibfield  {author} {\bibinfo {author} {\bibfnamefont {W.}~\bibnamefont {Lechner}}, \bibinfo {author} {\bibfnamefont {C.}~\bibnamefont {Dellago}}, \ and\ \bibinfo {author} {\bibfnamefont {P.~G.}\ \bibnamefont {Bolhuis}},\ }\bibfield  {title} {\enquote {\bibinfo {title} {Role of the {Prestructured} {Surface} {Cloud} in {Crystal} {Nucleation}},}\ }\href {\doibase 10.1103/PhysRevLett.106.085701} {\bibfield  {journal} {\bibinfo  {journal} {Phys. Rev. Lett.}\ }\textbf {\bibinfo {volume} {106}},\ \bibinfo {pages} {085701} (\bibinfo {year} {2011})}\BibitemShut {NoStop}%
\bibitem [{\citenamefont {Tan}, \citenamefont {Xu},\ and\ \citenamefont {Xu}(2014)}]{tan_visualizing_2014}%
  \BibitemOpen
  \bibfield  {author} {\bibinfo {author} {\bibfnamefont {P.}~\bibnamefont {Tan}}, \bibinfo {author} {\bibfnamefont {N.}~\bibnamefont {Xu}}, \ and\ \bibinfo {author} {\bibfnamefont {L.}~\bibnamefont {Xu}},\ }\bibfield  {title} {\enquote {\bibinfo {title} {Visualizing kinetic pathways of homogeneous nucleation in colloidal crystallization},}\ }\href {\doibase 10.1038/nphys2817} {\bibfield  {journal} {\bibinfo  {journal} {Nat. Phys.}\ }\textbf {\bibinfo {volume} {10}} (\bibinfo {year} {2014}),\ 10.1038/nphys2817}\BibitemShut {NoStop}%
\bibitem [{\citenamefont {Becker}\ \emph {et~al.}(2022)\citenamefont {Becker}, \citenamefont {Devijver}, \citenamefont {Molinier},\ and\ \citenamefont {Jakse}}]{becker_unsupervised_2022}%
  \BibitemOpen
  \bibfield  {author} {\bibinfo {author} {\bibfnamefont {S.}~\bibnamefont {Becker}}, \bibinfo {author} {\bibfnamefont {E.}~\bibnamefont {Devijver}}, \bibinfo {author} {\bibfnamefont {R.}~\bibnamefont {Molinier}}, \ and\ \bibinfo {author} {\bibfnamefont {N.}~\bibnamefont {Jakse}},\ }\bibfield  {title} {\enquote {\bibinfo {title} {Unsupervised topological learning approach of crystal nucleation},}\ }\href {\doibase 10.1038/s41598-022-06963-5} {\bibfield  {journal} {\bibinfo  {journal} {Sci Rep}\ }\textbf {\bibinfo {volume} {12}},\ \bibinfo {pages} {3195} (\bibinfo {year} {2022})}\BibitemShut {NoStop}%
\bibitem [{\citenamefont {Lechner}\ and\ \citenamefont {Dellago}(2008)}]{lechner_accurate_2008}%
  \BibitemOpen
  \bibfield  {author} {\bibinfo {author} {\bibfnamefont {W.}~\bibnamefont {Lechner}}\ and\ \bibinfo {author} {\bibfnamefont {C.}~\bibnamefont {Dellago}},\ }\bibfield  {title} {\enquote {\bibinfo {title} {Accurate determination of crystal structures based on averaged local bond order parameters},}\ }\href {\doibase 10.1063/1.2977970} {\bibfield  {journal} {\bibinfo  {journal} {J. Chem. Phys.}\ }\textbf {\bibinfo {volume} {129}},\ \bibinfo {pages} {114707} (\bibinfo {year} {2008})}\BibitemShut {NoStop}%
\bibitem [{\citenamefont {Chung}\ \emph {et~al.}(2022)\citenamefont {Chung}, \citenamefont {Freitas}, \citenamefont {Cheon},\ and\ \citenamefont {Reed}}]{chung_data-centric_2022}%
  \BibitemOpen
  \bibfield  {author} {\bibinfo {author} {\bibfnamefont {H.~W.}\ \bibnamefont {Chung}}, \bibinfo {author} {\bibfnamefont {R.}~\bibnamefont {Freitas}}, \bibinfo {author} {\bibfnamefont {G.}~\bibnamefont {Cheon}}, \ and\ \bibinfo {author} {\bibfnamefont {E.~J.}\ \bibnamefont {Reed}},\ }\bibfield  {title} {\enquote {\bibinfo {title} {Data-centric framework for crystal structure identification in atomistic simulations using machine learning},}\ }\href {\doibase 10.1103/PhysRevMaterials.6.043801} {\bibfield  {journal} {\bibinfo  {journal} {Phys. Rev. Mater.}\ }\textbf {\bibinfo {volume} {6}},\ \bibinfo {pages} {043801} (\bibinfo {year} {2022})}\BibitemShut {NoStop}%
\bibitem [{\citenamefont {Larsen}, \citenamefont {Schmidt},\ and\ \citenamefont {Schiøtz}(2016)}]{larsen_robust_2016}%
  \BibitemOpen
  \bibfield  {author} {\bibinfo {author} {\bibfnamefont {P.~M.}\ \bibnamefont {Larsen}}, \bibinfo {author} {\bibfnamefont {S.}~\bibnamefont {Schmidt}}, \ and\ \bibinfo {author} {\bibfnamefont {J.}~\bibnamefont {Schiøtz}},\ }\bibfield  {title} {{\enquote {\bibinfo {title} {Robust structural identification via polyhedral template matching},}\ }}\href {\doibase 10.1088/0965-0393/24/5/055007} {\bibfield  {journal} {\bibinfo  {journal} {Model. Simul. Mater. Sci. Eng.}\ }\textbf {\bibinfo {volume} {24}},\ \bibinfo {pages} {055007} (\bibinfo {year} {2016})}\BibitemShut {NoStop}%
\bibitem [{\citenamefont {Rahman}(1964)}]{rahman_correlations_1964}%
  \BibitemOpen
  \bibfield  {author} {\bibinfo {author} {\bibfnamefont {A.}~\bibnamefont {Rahman}},\ }\bibfield  {title} {\enquote {\bibinfo {title} {Correlations in the {Motion} of {Atoms} in {Liquid} {Argon}},}\ }\href {\doibase 10.1103/PhysRev.136.A405} {\bibfield  {journal} {\bibinfo  {journal} {Phys. Rev.}\ }\textbf {\bibinfo {volume} {136}},\ \bibinfo {pages} {A405--A411} (\bibinfo {year} {1964})}\BibitemShut {NoStop}%
\bibitem [{\citenamefont {Bulutoglu}\ \emph {et~al.}(2023)\citenamefont {Bulutoglu}, \citenamefont {Zalte}, \citenamefont {Nere}, \citenamefont {Ramkrishna},\ and\ \citenamefont {Corti}}]{bulutoglu_comprehensive_2023}%
  \BibitemOpen
  \bibfield  {author} {\bibinfo {author} {\bibfnamefont {P.~S.}\ \bibnamefont {Bulutoglu}}, \bibinfo {author} {\bibfnamefont {A.~S.}\ \bibnamefont {Zalte}}, \bibinfo {author} {\bibfnamefont {N.~K.}\ \bibnamefont {Nere}}, \bibinfo {author} {\bibfnamefont {D.}~\bibnamefont {Ramkrishna}}, \ and\ \bibinfo {author} {\bibfnamefont {D.~S.}\ \bibnamefont {Corti}},\ }\bibfield  {title} {\enquote {\bibinfo {title} {A comprehensive modeling approach for polymorph selection in {Lennard}-{Jones} crystallization},}\ }\href {\doibase 10.1063/5.0139476} {\bibfield  {journal} {\bibinfo  {journal} {J. Chem. Phys.}\ }\textbf {\bibinfo {volume} {158}},\ \bibinfo {pages} {134505} (\bibinfo {year} {2023})}\BibitemShut {NoStop}%
\bibitem [{\citenamefont {Plimpton}(1995)}]{plimpton_fast_1995}%
  \BibitemOpen
  \bibfield  {author} {\bibinfo {author} {\bibfnamefont {S.}~\bibnamefont {Plimpton}},\ }\bibfield  {title} {\enquote {\bibinfo {title} {Fast {Parallel} {Algorithms} for {Short}-{Range} {Molecular} {Dynamics}},}\ }\href {\doibase 10.1006/jcph.1995.1039} {\bibfield  {journal} {\bibinfo  {journal} {J. Comput. Phys.}\ }\textbf {\bibinfo {volume} {117}},\ \bibinfo {pages} {1--19} (\bibinfo {year} {1995})}\BibitemShut {NoStop}%
\bibitem [{\citenamefont {van~der Hoef}(2000)}]{van_der_hoef_free_2000}%
  \BibitemOpen
  \bibfield  {author} {\bibinfo {author} {\bibfnamefont {M.~A.}\ \bibnamefont {van~der Hoef}},\ }\bibfield  {title} {\enquote {\bibinfo {title} {Free energy of the {Lennard}-{Jones} solid},}\ }\href {\doibase 10.1063/1.1314342} {\bibfield  {journal} {\bibinfo  {journal} {J. Chem. Phys.}\ }\textbf {\bibinfo {volume} {113}},\ \bibinfo {pages} {8142--8148} (\bibinfo {year} {2000})}\BibitemShut {NoStop}%
\bibitem [{\citenamefont {Auer}\ and\ \citenamefont {Frenkel}(2001)}]{auer_prediction_2001}%
  \BibitemOpen
  \bibfield  {author} {\bibinfo {author} {\bibfnamefont {S.}~\bibnamefont {Auer}}\ and\ \bibinfo {author} {\bibfnamefont {D.}~\bibnamefont {Frenkel}},\ }\bibfield  {title} {\enquote {\bibinfo {title} {Prediction of absolute crystal-nucleation rate in hard-sphere colloids},}\ }\href {\doibase 10.1038/35059035} {\bibfield  {journal} {\bibinfo  {journal} {Nature}\ }\textbf {\bibinfo {volume} {409}},\ \bibinfo {pages} {1020--1023} (\bibinfo {year} {2001})}\BibitemShut {NoStop}%
\bibitem [{\citenamefont {Torrie}\ and\ \citenamefont {Valleau}(1977)}]{torrie_nonphysical_1977}%
  \BibitemOpen
  \bibfield  {author} {\bibinfo {author} {\bibfnamefont {G.~M.}\ \bibnamefont {Torrie}}\ and\ \bibinfo {author} {\bibfnamefont {J.~P.}\ \bibnamefont {Valleau}},\ }\bibfield  {title} {\enquote {\bibinfo {title} {Nonphysical sampling distributions in {Monte} {Carlo} free-energy estimation: {Umbrella} sampling},}\ }\href {\doibase 10.1016/0021-9991(77)90121-8} {\bibfield  {journal} {\bibinfo  {journal} {J. Comput. Phys.}\ }\textbf {\bibinfo {volume} {23}},\ \bibinfo {pages} {187--199} (\bibinfo {year} {1977})}\BibitemShut {NoStop}%
\bibitem [{\citenamefont {Laio}\ and\ \citenamefont {Parrinello}(2002)}]{laio_escaping_2002}%
  \BibitemOpen
  \bibfield  {author} {\bibinfo {author} {\bibfnamefont {A.}~\bibnamefont {Laio}}\ and\ \bibinfo {author} {\bibfnamefont {M.}~\bibnamefont {Parrinello}},\ }\bibfield  {title} {\enquote {\bibinfo {title} {Escaping free-energy minima},}\ }\href {\doibase 10.1073/pnas.202427399} {\bibfield  {journal} {\bibinfo  {journal} {Proc. Natl. Acad. Sci. U.S.A.}\ }\textbf {\bibinfo {volume} {99}},\ \bibinfo {pages} {12562--12566} (\bibinfo {year} {2002})}\BibitemShut {NoStop}%
\bibitem [{\citenamefont {Trudu}, \citenamefont {Donadio},\ and\ \citenamefont {Parrinello}(2006)}]{trudu_freezing_2006}%
  \BibitemOpen
  \bibfield  {author} {\bibinfo {author} {\bibfnamefont {F.}~\bibnamefont {Trudu}}, \bibinfo {author} {\bibfnamefont {D.}~\bibnamefont {Donadio}}, \ and\ \bibinfo {author} {\bibfnamefont {M.}~\bibnamefont {Parrinello}},\ }\bibfield  {title} {\enquote {\bibinfo {title} {Freezing of a {Lennard}-{Jones} {Fluid}: {From} {Nucleation} to {Spinodal} {Regime}},}\ }\href {\doibase 10.1103/PhysRevLett.97.105701} {\bibfield  {journal} {\bibinfo  {journal} {Phys. Rev. Lett.}\ }\textbf {\bibinfo {volume} {97}},\ \bibinfo {pages} {105701} (\bibinfo {year} {2006})}\BibitemShut {NoStop}%
\bibitem [{\citenamefont {Allen}, \citenamefont {Warren},\ and\ \citenamefont {ten Wolde}(2005)}]{allen_sampling_2005}%
  \BibitemOpen
  \bibfield  {author} {\bibinfo {author} {\bibfnamefont {R.~J.}\ \bibnamefont {Allen}}, \bibinfo {author} {\bibfnamefont {P.~B.}\ \bibnamefont {Warren}}, \ and\ \bibinfo {author} {\bibfnamefont {P.~R.}\ \bibnamefont {ten Wolde}},\ }\bibfield  {title} {\enquote {\bibinfo {title} {Sampling {Rare} {Switching} {Events} in {Biochemical} {Networks}},}\ }\href {\doibase 10.1103/PhysRevLett.94.018104} {\bibfield  {journal} {\bibinfo  {journal} {Phys. Rev. Lett.}\ }\textbf {\bibinfo {volume} {94}},\ \bibinfo {pages} {018104} (\bibinfo {year} {2005})}\BibitemShut {NoStop}%
\bibitem [{\citenamefont {Bolhuis}\ \emph {et~al.}(2002)\citenamefont {Bolhuis}, \citenamefont {Chandler}, \citenamefont {Dellago},\ and\ \citenamefont {Geissler}}]{bolhuis_transition_2002}%
  \BibitemOpen
  \bibfield  {author} {\bibinfo {author} {\bibfnamefont {P.~G.}\ \bibnamefont {Bolhuis}}, \bibinfo {author} {\bibfnamefont {D.}~\bibnamefont {Chandler}}, \bibinfo {author} {\bibfnamefont {C.}~\bibnamefont {Dellago}}, \ and\ \bibinfo {author} {\bibfnamefont {P.~L.}\ \bibnamefont {Geissler}},\ }\bibfield  {title} {\enquote {\bibinfo {title} {{TRANSITION} {PATH} {SAMPLING}: {Throwing} {Ropes} {Over} {Rough} {Mountain} {Passes}, in the {Dark}},}\ }\href {\doibase 10.1146/annurev.physchem.53.082301.113146} {\bibfield  {journal} {\bibinfo  {journal} {Annu. Rev. Phys. Chem.}\ }\textbf {\bibinfo {volume} {53}},\ \bibinfo {pages} {291--318} (\bibinfo {year} {2002})}\BibitemShut {NoStop}%
\bibitem [{\citenamefont {Peters}\ and\ \citenamefont {Trout}(2006)}]{peters_obtaining_2006}%
  \BibitemOpen
  \bibfield  {author} {\bibinfo {author} {\bibfnamefont {B.}~\bibnamefont {Peters}}\ and\ \bibinfo {author} {\bibfnamefont {B.~L.}\ \bibnamefont {Trout}},\ }\bibfield  {title} {\enquote {\bibinfo {title} {Obtaining reaction coordinates by likelihood maximization},}\ }\href {\doibase 10.1063/1.2234477} {\bibfield  {journal} {\bibinfo  {journal} {J. Chem. Phys.}\ }\textbf {\bibinfo {volume} {125}},\ \bibinfo {pages} {054108} (\bibinfo {year} {2006})}\BibitemShut {NoStop}%
\bibitem [{\citenamefont {Beckham}\ and\ \citenamefont {Peters}(2011)}]{beckham_optimizing_2011}%
  \BibitemOpen
  \bibfield  {author} {\bibinfo {author} {\bibfnamefont {G.~T.}\ \bibnamefont {Beckham}}\ and\ \bibinfo {author} {\bibfnamefont {B.}~\bibnamefont {Peters}},\ }\bibfield  {title} {\enquote {\bibinfo {title} {Optimizing {Nucleus} {Size} {Metrics} for {Liquid}–{Solid} {Nucleation} from {Transition} {Paths} of {Near}-{Nanosecond} {Duration}},}\ }\href {\doibase 10.1021/jz2002887} {\bibfield  {journal} {\bibinfo  {journal} {J. Phys. Chem. Lett.}\ }\textbf {\bibinfo {volume} {2}},\ \bibinfo {pages} {1133--1138} (\bibinfo {year} {2011})}\BibitemShut {NoStop}%
\bibitem [{\citenamefont {Steinhardt}, \citenamefont {Nelson},\ and\ \citenamefont {Ronchetti}(1983)}]{steinhardt_bond-orientational_1983}%
  \BibitemOpen
  \bibfield  {author} {\bibinfo {author} {\bibfnamefont {P.~J.}\ \bibnamefont {Steinhardt}}, \bibinfo {author} {\bibfnamefont {D.~R.}\ \bibnamefont {Nelson}}, \ and\ \bibinfo {author} {\bibfnamefont {M.}~\bibnamefont {Ronchetti}},\ }\bibfield  {title} {\enquote {\bibinfo {title} {Bond-orientational order in liquids and glasses},}\ }\href {\doibase 10.1103/PhysRevB.28.784} {\bibfield  {journal} {\bibinfo  {journal} {Phys. Rev. B}\ }\textbf {\bibinfo {volume} {28}},\ \bibinfo {pages} {784--805} (\bibinfo {year} {1983})}\BibitemShut {NoStop}%
\bibitem [{\citenamefont {van Meel}\ \emph {et~al.}(2012)\citenamefont {van Meel}, \citenamefont {Filion}, \citenamefont {Valeriani},\ and\ \citenamefont {Frenkel}}]{van_meel_parameter-free_2012}%
  \BibitemOpen
  \bibfield  {author} {\bibinfo {author} {\bibfnamefont {J.~A.}\ \bibnamefont {van Meel}}, \bibinfo {author} {\bibfnamefont {L.}~\bibnamefont {Filion}}, \bibinfo {author} {\bibfnamefont {C.}~\bibnamefont {Valeriani}}, \ and\ \bibinfo {author} {\bibfnamefont {D.}~\bibnamefont {Frenkel}},\ }\bibfield  {title} {\enquote {\bibinfo {title} {A parameter-free, solid-angle based, nearest-neighbor algorithm},}\ }\href {\doibase 10.1063/1.4729313} {\bibfield  {journal} {\bibinfo  {journal} {J. Chem. Phys.}\ }\textbf {\bibinfo {volume} {136}},\ \bibinfo {pages} {234107} (\bibinfo {year} {2012})}\BibitemShut {NoStop}%
\bibitem [{\citenamefont {ten Wolde}, \citenamefont {Ruiz‐Montero},\ and\ \citenamefont {Frenkel}(1996)}]{ten_wolde_numerical_1996}%
  \BibitemOpen
  \bibfield  {author} {\bibinfo {author} {\bibfnamefont {P.~R.}\ \bibnamefont {ten Wolde}}, \bibinfo {author} {\bibfnamefont {M.~J.}\ \bibnamefont {Ruiz‐Montero}}, \ and\ \bibinfo {author} {\bibfnamefont {D.}~\bibnamefont {Frenkel}},\ }\bibfield  {title} {\enquote {\bibinfo {title} {Numerical calculation of the rate of crystal nucleation in a {Lennard}‐{Jones} system at moderate undercooling},}\ }\href {\doibase 10.1063/1.471721} {\bibfield  {journal} {\bibinfo  {journal} {J. Chem. Phys.}\ }\textbf {\bibinfo {volume} {104}},\ \bibinfo {pages} {9932--9947} (\bibinfo {year} {1996})}\BibitemShut {NoStop}%
\bibitem [{\citenamefont {Espinosa}\ \emph {et~al.}(2016{\natexlab{b}})\citenamefont {Espinosa}, \citenamefont {Vega}, \citenamefont {Valeriani},\ and\ \citenamefont {Sanz}}]{espinosa_seeding_2016}%
  \BibitemOpen
  \bibfield  {author} {\bibinfo {author} {\bibfnamefont {J.~R.}\ \bibnamefont {Espinosa}}, \bibinfo {author} {\bibfnamefont {C.}~\bibnamefont {Vega}}, \bibinfo {author} {\bibfnamefont {C.}~\bibnamefont {Valeriani}}, \ and\ \bibinfo {author} {\bibfnamefont {E.}~\bibnamefont {Sanz}},\ }\bibfield  {title} {\enquote {\bibinfo {title} {Seeding approach to crystal nucleation},}\ }\href {\doibase 10.1063/1.4939641} {\bibfield  {journal} {\bibinfo  {journal} {J. Chem. Phys.}\ }\textbf {\bibinfo {volume} {144}},\ \bibinfo {pages} {034501} (\bibinfo {year} {2016}{\natexlab{b}})}\BibitemShut {NoStop}%
\bibitem [{\citenamefont {Geiger}\ and\ \citenamefont {Dellago}(2013)}]{geiger_neural_2013}%
  \BibitemOpen
  \bibfield  {author} {\bibinfo {author} {\bibfnamefont {P.}~\bibnamefont {Geiger}}\ and\ \bibinfo {author} {\bibfnamefont {C.}~\bibnamefont {Dellago}},\ }\bibfield  {title} {\enquote {\bibinfo {title} {Neural networks for local structure detection in polymorphic systems},}\ }\href {\doibase 10.1063/1.4825111} {\bibfield  {journal} {\bibinfo  {journal} {J. Chem. Phys.}\ }\textbf {\bibinfo {volume} {139}},\ \bibinfo {pages} {164105} (\bibinfo {year} {2013})}\BibitemShut {NoStop}%
\bibitem [{\citenamefont {Spellings}\ and\ \citenamefont {Glotzer}(2018)}]{spellings_machine_2018}%
  \BibitemOpen
  \bibfield  {author} {\bibinfo {author} {\bibfnamefont {M.}~\bibnamefont {Spellings}}\ and\ \bibinfo {author} {\bibfnamefont {S.~C.}\ \bibnamefont {Glotzer}},\ }\bibfield  {title} {\enquote {\bibinfo {title} {Machine learning for crystal identification and discovery},}\ }\href {\doibase https://doi.org/10.1002/aic.16157} {\bibfield  {journal} {\bibinfo  {journal} {AIChE}\ }\textbf {\bibinfo {volume} {64}},\ \bibinfo {pages} {2198--2206} (\bibinfo {year} {2018})}\BibitemShut {NoStop}%
\bibitem [{\citenamefont {Boattini}\ \emph {et~al.}(2018)\citenamefont {Boattini}, \citenamefont {Ram}, \citenamefont {Smallenburg},\ and\ \citenamefont {Filion}}]{boattini_neural-network-based_2018}%
  \BibitemOpen
  \bibfield  {author} {\bibinfo {author} {\bibfnamefont {E.}~\bibnamefont {Boattini}}, \bibinfo {author} {\bibfnamefont {M.}~\bibnamefont {Ram}}, \bibinfo {author} {\bibfnamefont {F.}~\bibnamefont {Smallenburg}}, \ and\ \bibinfo {author} {\bibfnamefont {L.}~\bibnamefont {Filion}},\ }\bibfield  {title} {\enquote {\bibinfo {title} {Neural-network-based order parameters for classification of binary hard-sphere crystal structures},}\ }\href {\doibase 10.1080/00268976.2018.1483537} {\bibfield  {journal} {\bibinfo  {journal} {Mol. Phys.}\ }\textbf {\bibinfo {volume} {116}},\ \bibinfo {pages} {3066--3075} (\bibinfo {year} {2018})}\BibitemShut {NoStop}%
\bibitem [{\citenamefont {S. DeFever}\ \emph {et~al.}(2019)\citenamefont {S. DeFever}, \citenamefont {Targonski}, \citenamefont {W. Hall}, \citenamefont {C. Smith},\ and\ \citenamefont {Sarupria}}]{sdefever_generalized_2019}%
  \BibitemOpen
  \bibfield  {author} {\bibinfo {author} {\bibfnamefont {R.}~\bibnamefont {S. DeFever}}, \bibinfo {author} {\bibfnamefont {C.}~\bibnamefont {Targonski}}, \bibinfo {author} {\bibfnamefont {S.}~\bibnamefont {W. Hall}}, \bibinfo {author} {\bibfnamefont {M.}~\bibnamefont {C. Smith}}, \ and\ \bibinfo {author} {\bibfnamefont {S.}~\bibnamefont {Sarupria}},\ }\bibfield  {title} {\enquote {\bibinfo {title} {A generalized deep learning approach for local structure identification in molecular simulations},}\ }\href {\doibase 10.1039/C9SC02097G} {\bibfield  {journal} {\bibinfo  {journal} {Chem. Sci.}\ }\textbf {\bibinfo {volume} {10}},\ \bibinfo {pages} {7503--7515} (\bibinfo {year} {2019})}\BibitemShut {NoStop}%
\bibitem [{\citenamefont {Adorf}\ \emph {et~al.}(2020)\citenamefont {Adorf}, \citenamefont {Moore}, \citenamefont {Melle},\ and\ \citenamefont {Glotzer}}]{adorf_analysis_2020}%
  \BibitemOpen
  \bibfield  {author} {\bibinfo {author} {\bibfnamefont {C.~S.}\ \bibnamefont {Adorf}}, \bibinfo {author} {\bibfnamefont {T.~C.}\ \bibnamefont {Moore}}, \bibinfo {author} {\bibfnamefont {Y.~J.~U.}\ \bibnamefont {Melle}}, \ and\ \bibinfo {author} {\bibfnamefont {S.~C.}\ \bibnamefont {Glotzer}},\ }\bibfield  {title} {\enquote {\bibinfo {title} {Analysis of {Self}-{Assembly} {Pathways} with {Unsupervised} {Machine} {Learning} {Algorithms}},}\ }\href {\doibase 10.1021/acs.jpcb.9b09621} {\bibfield  {journal} {\bibinfo  {journal} {J. Phys. Chem. B}\ }\textbf {\bibinfo {volume} {124}},\ \bibinfo {pages} {69--78} (\bibinfo {year} {2020})}\BibitemShut {NoStop}%
\bibitem [{\citenamefont {Coli}\ and\ \citenamefont {Dijkstra}(2021)}]{coli_artificial_2021}%
  \BibitemOpen
  \bibfield  {author} {\bibinfo {author} {\bibfnamefont {G.~M.}\ \bibnamefont {Coli}}\ and\ \bibinfo {author} {\bibfnamefont {M.}~\bibnamefont {Dijkstra}},\ }\bibfield  {title} {\enquote {\bibinfo {title} {An {Artificial} {Neural} {Network} {Reveals} the {Nucleation} {Mechanism} of a {Binary} {Colloidal} {AB13} {Crystal}},}\ }\href {\doibase 10.1021/acsnano.0c07541} {\bibfield  {journal} {\bibinfo  {journal} {ACS Nano}\ }\textbf {\bibinfo {volume} {15}},\ \bibinfo {pages} {4335--4346} (\bibinfo {year} {2021})}\BibitemShut {NoStop}%
\bibitem [{\citenamefont {Terao}(2023)}]{terao_anomaly_2023}%
  \BibitemOpen
  \bibfield  {author} {\bibinfo {author} {\bibfnamefont {T.}~\bibnamefont {Terao}},\ }\bibfield  {title} {\enquote {\bibinfo {title} {Anomaly detection for structural formation analysis by autoencoders: application to soft matters},}\ }\href {\doibase 10.1080/14786435.2023.2251408} {\bibfield  {journal} {\bibinfo  {journal} {Phil. Mag.}\ }\textbf {\bibinfo {volume} {103}},\ \bibinfo {pages} {2013--2028} (\bibinfo {year} {2023})}\BibitemShut {NoStop}%
\bibitem [{\citenamefont {Lee}\ \emph {et~al.}(2018)\citenamefont {Lee}, \citenamefont {Lee}, \citenamefont {Lee},\ and\ \citenamefont {Shin}}]{lee_simple_2018}%
  \BibitemOpen
  \bibfield  {author} {\bibinfo {author} {\bibfnamefont {K.}~\bibnamefont {Lee}}, \bibinfo {author} {\bibfnamefont {K.}~\bibnamefont {Lee}}, \bibinfo {author} {\bibfnamefont {H.}~\bibnamefont {Lee}}, \ and\ \bibinfo {author} {\bibfnamefont {J.}~\bibnamefont {Shin}},\ }\bibfield  {title} {\enquote {\bibinfo {title} {A {Simple} {Unified} {Framework} for {Detecting} {Out}-of-{Distribution} {Samples} and {Adversarial} {Attacks}},}\ }in\ \href {https://proceedings.neurips.cc/paper/2018/hash/abdeb6f575ac5c6676b747bca8d09cc2-Abstract.html} {\emph {\bibinfo {booktitle} {Adv. {Neural} {Inf}. {Process}. {Syst}.}}},\ Vol.~\bibinfo {volume} {31}\ (\bibinfo  {publisher} {Curran Associates, Inc.},\ \bibinfo {year} {2018})\BibitemShut {NoStop}%
\bibitem [{\citenamefont {Desgranges}\ and\ \citenamefont {Delhommelle}(2006)}]{desgranges_molecular_2006}%
  \BibitemOpen
  \bibfield  {author} {\bibinfo {author} {\bibfnamefont {C.}~\bibnamefont {Desgranges}}\ and\ \bibinfo {author} {\bibfnamefont {J.}~\bibnamefont {Delhommelle}},\ }\bibfield  {title} {\enquote {\bibinfo {title} {Molecular {Mechanism} for the {Cross}-{Nucleation} between {Polymorphs}},}\ }\href {\doibase 10.1021/ja063218f} {\bibfield  {journal} {\bibinfo  {journal} {J. Am. Chem. Soc.}\ }\textbf {\bibinfo {volume} {128}},\ \bibinfo {pages} {10368--10369} (\bibinfo {year} {2006})}\BibitemShut {NoStop}%
\bibitem [{\citenamefont {Eshet}, \citenamefont {Bruneval},\ and\ \citenamefont {Parrinello}(2008)}]{eshet_new_2008}%
  \BibitemOpen
  \bibfield  {author} {\bibinfo {author} {\bibfnamefont {H.}~\bibnamefont {Eshet}}, \bibinfo {author} {\bibfnamefont {F.}~\bibnamefont {Bruneval}}, \ and\ \bibinfo {author} {\bibfnamefont {M.}~\bibnamefont {Parrinello}},\ }\bibfield  {title} {\enquote {\bibinfo {title} {New {Lennard}-{Jones} metastable phase},}\ }\href {\doibase 10.1063/1.2953327} {\bibfield  {journal} {\bibinfo  {journal} {J. Chem. Phys.}\ }\textbf {\bibinfo {volume} {129}},\ \bibinfo {pages} {026101} (\bibinfo {year} {2008})}\BibitemShut {NoStop}%
\bibitem [{\citenamefont {Schwerdtfeger}\ and\ \citenamefont {Burrows}(2022)}]{schwerdtfeger_cuboidal_2022}%
  \BibitemOpen
  \bibfield  {author} {\bibinfo {author} {\bibfnamefont {P.}~\bibnamefont {Schwerdtfeger}}\ and\ \bibinfo {author} {\bibfnamefont {A.}~\bibnamefont {Burrows}},\ }\bibfield  {title} {\enquote {\bibinfo {title} {Cuboidal bcc to fcc {Transformation} of {Lennard}-{Jones} {Phases} under {High} {Pressure} {Derived} from {Exact} {Lattice} {Summations}},}\ }\href {\doibase 10.1021/acs.jpcc.2c01255} {\bibfield  {journal} {\bibinfo  {journal} {J. Phys. Chem. C}\ }\textbf {\bibinfo {volume} {126}},\ \bibinfo {pages} {8874--8882} (\bibinfo {year} {2022})}\BibitemShut {NoStop}%
\bibitem [{\citenamefont {Stukowski}(2012)}]{stukowski_structure_2012}%
  \BibitemOpen
  \bibfield  {author} {\bibinfo {author} {\bibfnamefont {A.}~\bibnamefont {Stukowski}},\ }\bibfield  {title} {\enquote {\bibinfo {title} {Structure identification methods for atomistic simulations of crystalline materials},}\ }\href {\doibase 10.1088/0965-0393/20/4/045021} {\bibfield  {journal} {\bibinfo  {journal} {Model. Simul. Mater. Sci. Eng.}\ }\textbf {\bibinfo {volume} {20}},\ \bibinfo {pages} {045021} (\bibinfo {year} {2012})}\BibitemShut {NoStop}%
\bibitem [{\citenamefont {Malins}\ \emph {et~al.}(2013)\citenamefont {Malins}, \citenamefont {Williams}, \citenamefont {Eggers},\ and\ \citenamefont {Royall}}]{malins_identification_2013}%
  \BibitemOpen
  \bibfield  {author} {\bibinfo {author} {\bibfnamefont {A.}~\bibnamefont {Malins}}, \bibinfo {author} {\bibfnamefont {S.~R.}\ \bibnamefont {Williams}}, \bibinfo {author} {\bibfnamefont {J.}~\bibnamefont {Eggers}}, \ and\ \bibinfo {author} {\bibfnamefont {C.~P.}\ \bibnamefont {Royall}},\ }\bibfield  {title} {\enquote {\bibinfo {title} {Identification of structure in condensed matter with the topological cluster classification},}\ }\href {\doibase 10.1063/1.4832897} {\bibfield  {journal} {\bibinfo  {journal} {J. Chem. Phys.}\ }\textbf {\bibinfo {volume} {139}},\ \bibinfo {pages} {234506} (\bibinfo {year} {2013})}\BibitemShut {NoStop}%
\bibitem [{\citenamefont {Swope}\ and\ \citenamefont {Andersen}(1990)}]{swope_106-particle_1990}%
  \BibitemOpen
  \bibfield  {author} {\bibinfo {author} {\bibfnamefont {W.~C.}\ \bibnamefont {Swope}}\ and\ \bibinfo {author} {\bibfnamefont {H.~C.}\ \bibnamefont {Andersen}},\ }\bibfield  {title} {\enquote {\bibinfo {title} {\$\{10\}{\textasciicircum}\{6\}\$-particle molecular-dynamics study of homogeneous nucleation of crystals in a supercooled atomic liquid},}\ }\href {\doibase 10.1103/PhysRevB.41.7042} {\bibfield  {journal} {\bibinfo  {journal} {Phys. Rev. B}\ }\textbf {\bibinfo {volume} {41}},\ \bibinfo {pages} {7042--7054} (\bibinfo {year} {1990})}\BibitemShut {NoStop}%
\bibitem [{\citenamefont {Menon}\ \emph {et~al.}(2020)\citenamefont {Menon}, \citenamefont {Díaz~Leines}, \citenamefont {Drautz},\ and\ \citenamefont {Rogal}}]{menon_role_2020}%
  \BibitemOpen
  \bibfield  {author} {\bibinfo {author} {\bibfnamefont {S.}~\bibnamefont {Menon}}, \bibinfo {author} {\bibfnamefont {G.}~\bibnamefont {Díaz~Leines}}, \bibinfo {author} {\bibfnamefont {R.}~\bibnamefont {Drautz}}, \ and\ \bibinfo {author} {\bibfnamefont {J.}~\bibnamefont {Rogal}},\ }\bibfield  {title} {\enquote {\bibinfo {title} {Role of pre-ordered liquid in the selection mechanism of crystal polymorphs during nucleation},}\ }\href {\doibase 10.1063/5.0017575} {\bibfield  {journal} {\bibinfo  {journal} {J. Chem. Phys.}\ }\textbf {\bibinfo {volume} {153}},\ \bibinfo {pages} {104508} (\bibinfo {year} {2020})}\BibitemShut {NoStop}%
\bibitem [{\citenamefont {Díaz~Leines}\ and\ \citenamefont {Rogal}(2018)}]{diaz_leines_maximum_2018}%
  \BibitemOpen
  \bibfield  {author} {\bibinfo {author} {\bibfnamefont {G.}~\bibnamefont {Díaz~Leines}}\ and\ \bibinfo {author} {\bibfnamefont {J.}~\bibnamefont {Rogal}},\ }\bibfield  {title} {\enquote {\bibinfo {title} {Maximum {Likelihood} {Analysis} of {Reaction} {Coordinates} during {Solidification} in {Ni}},}\ }\href {\doibase 10.1021/acs.jpcb.8b08718} {\bibfield  {journal} {\bibinfo  {journal} {J. Phys. Chem. B}\ }\textbf {\bibinfo {volume} {122}},\ \bibinfo {pages} {10934--10942} (\bibinfo {year} {2018})}\BibitemShut {NoStop}%
\bibitem [{\citenamefont {Gispen}\ \emph {et~al.}(2023)\citenamefont {Gispen}, \citenamefont {Coli}, \citenamefont {van Damme}, \citenamefont {Royall},\ and\ \citenamefont {Dijkstra}}]{gispen_crystal_2023}%
  \BibitemOpen
  \bibfield  {author} {\bibinfo {author} {\bibfnamefont {W.}~\bibnamefont {Gispen}}, \bibinfo {author} {\bibfnamefont {G.~M.}\ \bibnamefont {Coli}}, \bibinfo {author} {\bibfnamefont {R.}~\bibnamefont {van Damme}}, \bibinfo {author} {\bibfnamefont {C.~P.}\ \bibnamefont {Royall}}, \ and\ \bibinfo {author} {\bibfnamefont {M.}~\bibnamefont {Dijkstra}},\ }\bibfield  {title} {\enquote {\bibinfo {title} {Crystal {Polymorph} {Selection} {Mechanism} of {Hard} {Spheres} {Hidden} in the {Fluid}},}\ }\href {\doibase https://doi.org/10.1021/acsnano.3c02182} {\bibfield  {journal} {\bibinfo  {journal} {ACS Nano}\ }\textbf {\bibinfo {volume} {17}},\ \bibinfo {pages} {8807--8814} (\bibinfo {year} {2023})}\BibitemShut {NoStop}%
\bibitem [{\citenamefont {Díaz~Leines}, \citenamefont {Drautz},\ and\ \citenamefont {Rogal}(2017)}]{diaz_leines_atomistic_2017}%
  \BibitemOpen
  \bibfield  {author} {\bibinfo {author} {\bibfnamefont {G.}~\bibnamefont {Díaz~Leines}}, \bibinfo {author} {\bibfnamefont {R.}~\bibnamefont {Drautz}}, \ and\ \bibinfo {author} {\bibfnamefont {J.}~\bibnamefont {Rogal}},\ }\bibfield  {title} {\enquote {\bibinfo {title} {Atomistic insight into the non-classical nucleation mechanism during solidification in {Ni}},}\ }\href {\doibase 10.1063/1.4980082} {\bibfield  {journal} {\bibinfo  {journal} {J. Chem. Phys.}\ }\textbf {\bibinfo {volume} {146}},\ \bibinfo {pages} {154702} (\bibinfo {year} {2017})}\BibitemShut {NoStop}%
\bibitem [{\citenamefont {Schöll-Paschinger}\ and\ \citenamefont {Dellago}(2010)}]{scholl-paschinger_demixing_2010}%
  \BibitemOpen
  \bibfield  {author} {\bibinfo {author} {\bibfnamefont {E.}~\bibnamefont {Schöll-Paschinger}}\ and\ \bibinfo {author} {\bibfnamefont {C.}~\bibnamefont {Dellago}},\ }\bibfield  {title} {\enquote {\bibinfo {title} {Demixing of a binary symmetric mixture studied with transition path sampling},}\ }\href {\doibase 10.1063/1.3486173} {\bibfield  {journal} {\bibinfo  {journal} {J. Chem. Phys.}\ }\textbf {\bibinfo {volume} {133}},\ \bibinfo {pages} {104505} (\bibinfo {year} {2010})}\BibitemShut {NoStop}%
\bibitem [{\citenamefont {Falkner}\ \emph {et~al.}(2024)\citenamefont {Falkner}, \citenamefont {Coretti}, \citenamefont {Peters}, \citenamefont {Bolhuis},\ and\ \citenamefont {Dellago}}]{falkner_revisiting_2024}%
  \BibitemOpen
  \bibfield  {author} {\bibinfo {author} {\bibfnamefont {S.}~\bibnamefont {Falkner}}, \bibinfo {author} {\bibfnamefont {A.}~\bibnamefont {Coretti}}, \bibinfo {author} {\bibfnamefont {B.}~\bibnamefont {Peters}}, \bibinfo {author} {\bibfnamefont {P.~G.}\ \bibnamefont {Bolhuis}}, \ and\ \bibinfo {author} {\bibfnamefont {C.}~\bibnamefont {Dellago}},\ }\href {\doibase 10.48550/arXiv.2408.03054} {\enquote {\bibinfo {title} {Revisiting {Shooting} {Point} {Monte} {Carlo} {Methods} for {Transition} {Path} {Sampling}},}\ } (\bibinfo {year} {2024}),\ \bibinfo {note} {arXiv:2408.03054 [cond-mat, physics:physics]}\BibitemShut {NoStop}%
\bibitem [{\citenamefont {Ramasubramani}\ \emph {et~al.}(2020)\citenamefont {Ramasubramani}, \citenamefont {Dice}, \citenamefont {Harper}, \citenamefont {Spellings}, \citenamefont {Anderson},\ and\ \citenamefont {Glotzer}}]{ramasubramani_freud_2020}%
  \BibitemOpen
  \bibfield  {author} {\bibinfo {author} {\bibfnamefont {V.}~\bibnamefont {Ramasubramani}}, \bibinfo {author} {\bibfnamefont {B.~D.}\ \bibnamefont {Dice}}, \bibinfo {author} {\bibfnamefont {E.~S.}\ \bibnamefont {Harper}}, \bibinfo {author} {\bibfnamefont {M.~P.}\ \bibnamefont {Spellings}}, \bibinfo {author} {\bibfnamefont {J.~A.}\ \bibnamefont {Anderson}}, \ and\ \bibinfo {author} {\bibfnamefont {S.~C.}\ \bibnamefont {Glotzer}},\ }\bibfield  {title} {\enquote {\bibinfo {title} {freud: {A} software suite for high throughput analysis of particle simulation data},}\ }\href {\doibase 10.1016/j.cpc.2020.107275} {\bibfield  {journal} {\bibinfo  {journal} {Comput. Phys. Commun.}\ }\textbf {\bibinfo {volume} {254}},\ \bibinfo {pages} {107275} (\bibinfo {year} {2020})}\BibitemShut {NoStop}%
\bibitem [{\citenamefont {Filion}\ \emph {et~al.}(2010)\citenamefont {Filion}, \citenamefont {Hermes}, \citenamefont {Ni},\ and\ \citenamefont {Dijkstra}}]{filion_crystal_2010}%
  \BibitemOpen
  \bibfield  {author} {\bibinfo {author} {\bibfnamefont {L.}~\bibnamefont {Filion}}, \bibinfo {author} {\bibfnamefont {M.}~\bibnamefont {Hermes}}, \bibinfo {author} {\bibfnamefont {R.}~\bibnamefont {Ni}}, \ and\ \bibinfo {author} {\bibfnamefont {M.}~\bibnamefont {Dijkstra}},\ }\bibfield  {title} {\enquote {\bibinfo {title} {Crystal nucleation of hard spheres using molecular dynamics, umbrella sampling, and forward flux sampling: {A} comparison of simulation techniques},}\ }\href {\doibase 10.1063/1.3506838} {\bibfield  {journal} {\bibinfo  {journal} {J. Chem. Phys.}\ }\textbf {\bibinfo {volume} {133}},\ \bibinfo {pages} {244115} (\bibinfo {year} {2010})}\BibitemShut {NoStop}%
\bibitem [{\citenamefont {Pedregosa}\ \emph {et~al.}(2011)\citenamefont {Pedregosa}, \citenamefont {Varoquaux}, \citenamefont {Gramfort}, \citenamefont {Michel}, \citenamefont {Thirion}, \citenamefont {Grisel}, \citenamefont {Blondel}, \citenamefont {Prettenhofer}, \citenamefont {Weiss}, \citenamefont {Dubourg}, \citenamefont {Vanderplas}, \citenamefont {Passos}, \citenamefont {Cournapeau}, \citenamefont {Brucher}, \citenamefont {Perrot},\ and\ \citenamefont {Duchesnay}}]{scikit-learn}%
  \BibitemOpen
  \bibfield  {author} {\bibinfo {author} {\bibfnamefont {F.}~\bibnamefont {Pedregosa}}, \bibinfo {author} {\bibfnamefont {G.}~\bibnamefont {Varoquaux}}, \bibinfo {author} {\bibfnamefont {A.}~\bibnamefont {Gramfort}}, \bibinfo {author} {\bibfnamefont {V.}~\bibnamefont {Michel}}, \bibinfo {author} {\bibfnamefont {B.}~\bibnamefont {Thirion}}, \bibinfo {author} {\bibfnamefont {O.}~\bibnamefont {Grisel}}, \bibinfo {author} {\bibfnamefont {M.}~\bibnamefont {Blondel}}, \bibinfo {author} {\bibfnamefont {P.}~\bibnamefont {Prettenhofer}}, \bibinfo {author} {\bibfnamefont {R.}~\bibnamefont {Weiss}}, \bibinfo {author} {\bibfnamefont {V.}~\bibnamefont {Dubourg}}, \bibinfo {author} {\bibfnamefont {J.}~\bibnamefont {Vanderplas}}, \bibinfo {author} {\bibfnamefont {A.}~\bibnamefont {Passos}}, \bibinfo {author} {\bibfnamefont {D.}~\bibnamefont {Cournapeau}}, \bibinfo {author} {\bibfnamefont {M.}~\bibnamefont {Brucher}}, \bibinfo {author} {\bibfnamefont {M.}~\bibnamefont {Perrot}}, \ and\ \bibinfo {author} {\bibfnamefont
  {E.}~\bibnamefont {Duchesnay}},\ }\bibfield  {title} {\enquote {\bibinfo {title} {Scikit-learn: Machine learning in {P}ython},}\ }\href@noop {} {\bibfield  {journal} {\bibinfo  {journal} {J. Mach. Learn. Res.}\ }\textbf {\bibinfo {volume} {12}},\ \bibinfo {pages} {2825--2830} (\bibinfo {year} {2011})}\BibitemShut {NoStop}%
\bibitem [{\citenamefont {Breiman}(2001)}]{breiman_random_2001}%
  \BibitemOpen
  \bibfield  {author} {\bibinfo {author} {\bibfnamefont {L.}~\bibnamefont {Breiman}},\ }\bibfield  {title} {\enquote {\bibinfo {title} {Random {Forests}},}\ }\href {\doibase 10.1023/A:1010933404324} {\bibfield  {journal} {\bibinfo  {journal} {Mach. Learn.}\ }\textbf {\bibinfo {volume} {45}},\ \bibinfo {pages} {5--32} (\bibinfo {year} {2001})}\BibitemShut {NoStop}%
\end{thebibliography}%

\end{document}